\documentclass[sigconf]{acmart}
\acmConference[]{}{}{}
\setcopyright{none}

\usepackage{balance}  
\usepackage{color,listings,booktabs,graphicx}
\usepackage{soul}
\usepackage{caption,subcaption}
\usepackage{arydshln} 
\usepackage{multirow}
\usepackage{pgfplots}
\usepackage{filecontents}
\usepackage{xspace}
\usepackage{enumitem} 
\setlist{noitemsep,topsep=2pt,parsep=2pt,partopsep=0pt,leftmargin=*} 
\renewcommand{\paragraph}[1]{\vspace{3pt} \noindent {\bf #1.}} 
\newtheorem{myexample}{Example}

\definecolor{darkgreen}{RGB}{0,128,0}
\definecolor{dkpink}{RGB}{200,0,100}
\definecolor{gray}{RGB}{128,128,128}

\usepackage{amssymb}
\usepackage{pifont}
\newcommand{\tdx}{\ding{55}} 

\def\SC{SC\xspace}
\def\systemName{Beta\xspace}
\def\IODS{IODS\xspace}

\lstdefinelanguage{Scala}{
	keywords={abstract,case,catch,char,class,def,do,else,extends,false,final,finally,for,if,implicit,import,%
	match,mixin,new,null,object,override,package,lazy,private,protected,requires,return,sealed,super,this,%
	throw,trait,true,try,type,val,var,while,with,yield},
	otherkeywords={=>,<-,<\%,<:,>:,\#,@},sensitive=true,
	morecomment=[l]{//},	morecomment=[n]{/*}{*/},
	morestring=[b]",morestring=[b]',morestring=[b]""",showstringspaces=false
}

\lstdefinelanguage{ScalaXact}{
	keywords={abstract,case,catch,class,def,do,else,extends,false,final,finally,for,if,implicit,import,%
	match,mixin,new,null,object,override,package,lazy,private,protected,requires,return,sealed,super,this,%
	throw,trait,true,try,type,val,var,while,with,yield,SELECT,FROM,WHERE,EXEC,ORDER,BY,DELETE,SQL,COMMIT,WORK},
	otherkeywords={=>,<-,<\%,<:,>:,\#,@},sensitive=true,
	morecomment=[l]{//},	morecomment=[n]{/*}{*/},
	morestring=[b]",morestring=[b]',morestring=[b]""",showstringspaces=false
}

\newenvironment{paperexample}{
   \begin{myexample} \em }{
   \end{myexample}}
\newenvironment{contexample}{
   \addtocounter{myexample}{-1} \begin{myexample}[continued] \em}{
   \end{myexample}}

\title{Compiling Database Application Programs}

\renewcommand{\shorttitle}{Compiling Database Application Programs}

\author{
Mohammad Dashti, 
Sachin Basil John, 
Thierry Coppey, \\
Amir Shaikhha, 
Vojin Jovanovic, 
and Christoph Koch\\
\hspace{1cm}\\
EPFL DATA Lab\hspace{0.3cm} \{firstname\}.\{lastname\}@epfl.ch\\
}
\gdef\shortauthors{EPFL DATA Lab}

\begin{document}

\begin{abstract}
There is a trend towards increased specialization of data management software for performance reasons.  In this paper, we study the automatic specialization and optimization of database application programs -- sequences of queries and updates, augmented with control flow constructs as they appear in database scripts, UDFs, transactional workloads and triggers in languages such as PL/SQL. We show how to build an optimizing compiler for database application programs using generative programming and state-of-the-art compiler technology.

We evaluate a hand-optimized low-level implementation of TPC-C, and identify the key optimization techniques that account for its good performance. Our compiler fully automates these optimizations and, applied to this benchmark, outperforms the manually optimized baseline by a factor of two. By selectively disabling some of the optimizations in the compiler, we derive a clinical and precise way of obtaining insight into their individual performance contributions.
\end{abstract}

\maketitle

\section{Introduction} \label{sec:intro}

Can good low-level code for database application programs be automatically generated?
Does it regularly require human creativity or is there a methodology that can be automated and systematically applied?
Is the day-to-day job of systems programmers fundamentally highly creative or could it be considered routine in the future?

Application development is preferable in high-level languages due to many reasons. One common reason is to have a faster development lifecycle due to the higher level of abstraction that they provide. The first prototype of a data processing application is usually written in a high-level language, which is then used to evaluate the ability of the program to produce the desired results. Then, if the prototype achieves the performance goals required by its users, it can be used in the production environment. Otherwise, the code needs to be further optimized for performance.

The task of optimizing an application program can, in principle, either be done manually by an expert programmer or by an automated tool. Code optimization by an expert programmer mainly takes two forms. The first is to, wherever possible, replace an algorithm or local implementation choice by another one that is more efficient in that particular scenario (specialization). The second is to translate parts of the program to a lower abstraction level (by using a lower-level language or inlining library calls) to create opportunities for further optimizations, resulting in a tighter control of the underlying machine through the program code.

Manual optimization can produce a more efficient system, but it is work-intensive and thus expensive.
Code lowering and specialization dramatically increase the maintenance cost of the system due to a substantial increase in the size and complexity of the code. If done by a human, there is no guarantee that all the choices taken by the systems programmer are optimal or even improvements, nor can one rely on a human to apply a given known set of optimization techniques consistently and thoroughly -- thus there is no optimality guarantee of any form.

The alternative way of achieving code specialization and optimization is to have it be done {\em automatically} by a system that satisfies the broad label of compiler or code synthesizer.
We are not referring to the code specialization tasks we have, over the past decades, come to universally trust our mainstream compilers to do reliably, but tasks that currently are performed by systems programmers, even when they use a mainstream compiler. There is currently a good deal of excitement in research circles, mostly revolving around domain-specific compiler technology~\cite{hekaton,legobase,neumann_llvm,sc,delitejournal}, that suggests that automation of the systems programmer in a domain such as databases that may soon become reality.

In this paper, we study the compilation of database application programs and propose how to build a compiler-based infrastructure for whole-system specialization. 
Ideally, using our compiler, database application programmers may use high-level programming abstractions without worrying about performance degradation ({\em abstraction without regret} as in \cite{KochManifesto}). The compiler simulates the work of a human systems programmer and employs similar methods and optimizations. The thesis is that the resulting code can match or even outperform the code created by human experts.

For 40 years, since the earliest times of relational DBMS, the compilation of query plans has been studied.\footnote{In the very beginning of the System R project, query execution was based on compilation.
Because the 1970s compiler technology was inconvenient in a rapidly progressing project, this compiler was replaced by the now standard interpretation approach just before the first release.
Also many new ideas had to be explored: it was much faster to implement an algorithm directly than to write a second algorithm that outputs the code of the first \cite{DBLP:journals/cacm/ChamberlinABGKLLMPPSSSTWY81}.} Recently, we have seen a revival of query compilation in contexts such as stream processing systems (IBM Spade and Streambase) and analytical queries (DBToaster \cite{dbt_delta}, LegoBase \cite{legobase, legobase_tods} and Hyper \cite{neumann_llvm}).
However,  the optimized compilation of application and transaction code is largely unknown territory, only recently entered by Microsoft's Hekaton system \cite {hekaton}, for a limited subset of T-SQL, the domain-specific language (DSL) of Microsoft SQL Server for writing transaction programs.
Good speed\-ups compared to classical interpreted evaluation have been reported,
but it is still open whether such a compiler can rival the performance hand-optimized code can achieve (cf.\ \cite{hstore_rewrite}) and what is the approximated individual impact of these optimizations.

In contrast to queries, which are phrased in a limited language, database application programs have more variability and mix queries and updates with general-purpose code. Therefore, an application compiler's internal code representations need to be more generic than those of a query compiler. But once such a powerful language is supported, the system can compile, inline and optimize the application programs {\em together} with generic server code, achieving a maximal degree of specialization and thus performance.

While modern compilers for general-purpose programming languages perform
many generic optimizations such as inlining, fusion and deforestation, they do so conservatively. Obtaining maximal performance necessitates moving away from conservative heuristics and pessimistic approaches of the general purpose compilers. In addition, there exist some other optimizations that are in general impossible or meaningless
outside the domain of database applications. For example, general purpose compilers treat all function calls equally, even though we could leverage \emph{domain-specific knowledge} (e.g., that joins are commutative). 
Since these compilers lack domain/data-structure specific optimizations,
they are not able to match the performance
of code written in a low-level language by a human expert.

\def\tdl{\vspace{2pt}\\ \hdashline[1pt/2pt] \vspace{-8pt}\\} 
\def\tdm{\vspace{1pt}\\\cdashline{2-9}[1pt/2pt]\vspace{-9pt}\\} 
\def\tdc{$\checkmark$} 
\def\tdo#1{\rotatebox{90}{#1}} 

\begin{table}
\vspace{6mm}
\caption{Optimizations employed in an expertly hand-written implementation of TPC-C \protect\cite{hstore_rewrite}, and by \systemName.} 
\vspace{-2mm}
\begin{center}\small\begin{tabular}
    {l@{\hskip 8pt}l@{\hskip 8pt}c@{\hskip 6pt}c@{\hskip 6pt}c@{\hskip 4pt}c@{\hskip 6pt}c@{\hskip 6pt}c@{\hskip 6pt}c}
    \begin{minipage}[b]{1.7cm}\bf TPC-C\\transaction\end{minipage}
    & \bf Version & \tdo{Mutable}\,\tdo{records} & \tdo{Inlining} & \tdo{Removing}\,\tdo{index update} & \tdo{Data structure}\,\tdo{specialization} & \tdo{Index}\,\tdo{introduction} & \tdo{Hoisting}\,\tdo{reusable objects}  & \tdo{Using}\,\tdo{runtime info}
\\ \midrule
\multirow{2}{*}{NewOrder}
    & \systemName    & \tdc & \tdc & \tdc & \tdc & \tdc & \tdc & \tdc  \tdm
    & Hand-written            & \tdc & \tdx & \tdc & \tdc & \tdc & \tdx & \tdc  \tdl
\multirow{2}{*}{Payment}
    & \systemName    & \tdc & \tdc & \tdc & \tdc & \tdc & \tdc & \tdc  \tdm
    & Hand-written            & \tdc & \tdx & \tdc & \tdc & \tdx & \tdx & \tdc  \tdl
\multirow{2}{*}{OrderStatus}
    & \systemName    &  \tdc & \tdc & -- & \tdc & \tdc & \tdc & \tdc  \tdm
    & Hand-written             &  \tdc & \tdx & -- & \tdc & \tdx & \tdx & \tdc  \tdl
\multirow{2}{*}{Delivery}
    & \systemName    & \tdc & \tdc & \tdc & \tdc & \tdc & \tdc & \tdc \tdm
    & Hand-written            & \tdc & \tdx & \tdc & \tdc & \tdx & \tdx & \tdc  \tdl
\multirow{2}{*}{StockLevel}
    & \systemName    &  \tdc & \tdc & -- & \tdc & \tdc & \tdc & \tdc \tdm
    & Hand-written            &  \tdc & \tdx & -- & \tdc & \tdx & \tdx & \tdc  \\ \bottomrule \vspace{-6pt}
\end{tabular}
\end{center}

\vspace{-7mm}\label{tab:opts_used}
\end{table}

Although database application programs can use the control structures of general-purpose programming languages, for the purpose of optimization, the task is greatly simplified -- and the automatic generation of expertly optimized code becomes
feasible -- by the fact that the language used remains domain-specific by its use of a very narrow and known set of data structures.
Database application programs tend to use the database's storage structures (tables, or the database connectivity library's proxies -- ``result sets'') as their only dynamic data structures. Creating special compiler optimizations related to such collection data structures can thus be highly effective, particularly if one keeps in mind that domain-specific (abstract) data types are a natural focal point for most domain-specific compiler optimizations.

\smallskip

\noindent
{\bf \systemName}.
We propose a database application compilation suite called \systemName\footnote{Referring to beta-reduction in lambda calculus, which \em{compiles applications}.}; its novelty lies in its adoption of modern compiler ideas such as {\em staging} \cite{DBLP:journals/cacm/ChamberlinABGKLLMPPSSSTWY81, neumann_llvm}. Optimizations are applied on the input program in multiple stages, and in each stage, parts of the program are partially evaluated based on the arguments known at compile time. \systemName also has a runtime engine that is flexible enough to run the generated code in any combination of optimizations. Moreover, the runtime engine exposes the internals of the data structures to \systemName, and the latter compiles them together with the application programs.

{\bf Inside-Out Data Structures}. Data structures designed to be used by programmers usually follow the \emph{information hiding principle}, which means they do not expose their internal implementation to programmers.
Even though a compiler does not have to follow the information hiding principle, the analysis for such an optimization (to disregard the additional programming layers created only for applying the information hiding principle) is almost impossible for a generic compiler. However, \systemName can perform such an optimization because it can reason more easily about the behavior of the limited set of  data structures that are used. More details about the design of such an execution engine are described in section \ref{sec:datastructures}.

In this paper, we first focus on achieving the best possible performance in a main-memory, single-partition, single-core case where programs are short-running and run in sequence. However, many database application programs run concurrently and in parallel to make the best use of resources available in the system. In such an environment, the concurrency control mechanism used in the execution engine of the database system plays an important role in its performance. It also impacts how a database application compiler, such as \systemName, is employed. We later discuss the concurrency aspects in section~\ref{sec:concurrency}. Moreover, to provide durability, \systemName makes use of command-logging similar to H-Store \cite{jones10}, but is extensible to accept other mechanisms as well.

{\bf Beating the Humans}. \systemName is inspired by the highly optimized manual implementation of TPC-C that was presented in \cite{hstore_rewrite}, and which was used there to motivate the creation of the H-Store system (as the implementation is shown to outperform existing OLTP systems by two orders of magnitude). We use this codebase, developed by a group of acknowledged experts, to make headway on the questions we asked at the start of this section. \systemName seems to be the natural completion of the H-Store vision: Writing low-level code that achieves the performance called for is very hard and there are few human programmers able to achieve it. Our focus on this use case is also due to the absence of any second such highly optimized low-level codebase in this space that we are aware of.

We acknowledge that the compilation of database application programs is a very broad space, touching many rich fields such as concurrency control and durability, which cannot all be satisfactorily addressed here.
Moreover, the challenge can be addressed in fundamentally opposing ways depending on choices such as whether the application code is to be run on the client or the server.
We cannot -- and will not claim to -- offer a design to address all these directions. Instead, we will derive insights from one well-known case study, and see to what extent we can draw generalizable and relevant conclusions from it. The compiler we have created is a work in-progress towards an OLTP system and is meant to generalize these insights (and we evaluate it on a second use case in this paper), but we do not claim that it provides a solution to all questions related to compiling database application code. In some cases, we make design choices specifically to preserve the case study from \cite{hstore_rewrite} as a valid baseline for experimentation.

In our experiments, we show that the compiler, with its superior stamina, even beats the human experts at their own game, making the vision of \cite{hstore_rewrite} come true: {\em a database system with a high-level language performs 100 times faster than a classical RDBMS}, and is competitive with ``ideal'' hand-optimized code.
We demonstrate this for the TPC-C benchmark, evaluating \systemName
against hand-optimized codebases. Our results show that the TPC-C programs compiled using \systemName are twice as efficient as the hand-optimized implementation of \cite{hstore_rewrite}.
In Table~\ref{tab:opts_used} we compare, in terms of key optimizations employed,
this hand-optimized implementation with the code generated by \systemName. Some optimizations do not apply to the TPC-C transaction programs, but our
compiler uses all the optimization ideas employed by the human expert and applies them more thoroughly. The details of these optimizations are discussed in section~\ref{sec:optimizations}.

To get further insights into the generality of our technique, we experiment with incremental view maintenance (IVM) trigger programs. DBToaster \cite{dbt_delta} generates IVM engines in two steps: SQL queries are transformed into trigger programs (front-end), which are then converted into executable code (back-end). By replacing the back-end with \systemName, we gained two orders of magnitude performance improvement over the original DBToaster compiler, which does not perform such extensive optimizations.

\smallskip
In summary, the contributions of this paper are as follows:

\begin{itemize}
\item \systemName, a database application execution infrastructure with a staging compiler in its heart.
It has an extended DSL that provides a minimal set of features required by database application programmers, on top of which well-defined optimizations are applied. It applies these optimizations more thoroughly and generates code that outperforms the hand-written code by human experts. 
\item A mechanism for the core supporting data structures to expose their internal functioning to the compiler.
\item Identifying key optimizations that are relevant in database applications and detailed experiments quantifying their impact.
\end{itemize}
\smallskip
We examine the architecture of \systemName in detail in the next section, and the structure of the rest of this paper closely follows the above order of contributions.

\section{Architecture}\label{sec:arch}

\begin{figure}[t!]
\begin{center}
\leavevmode
\includegraphics[width=\columnwidth]{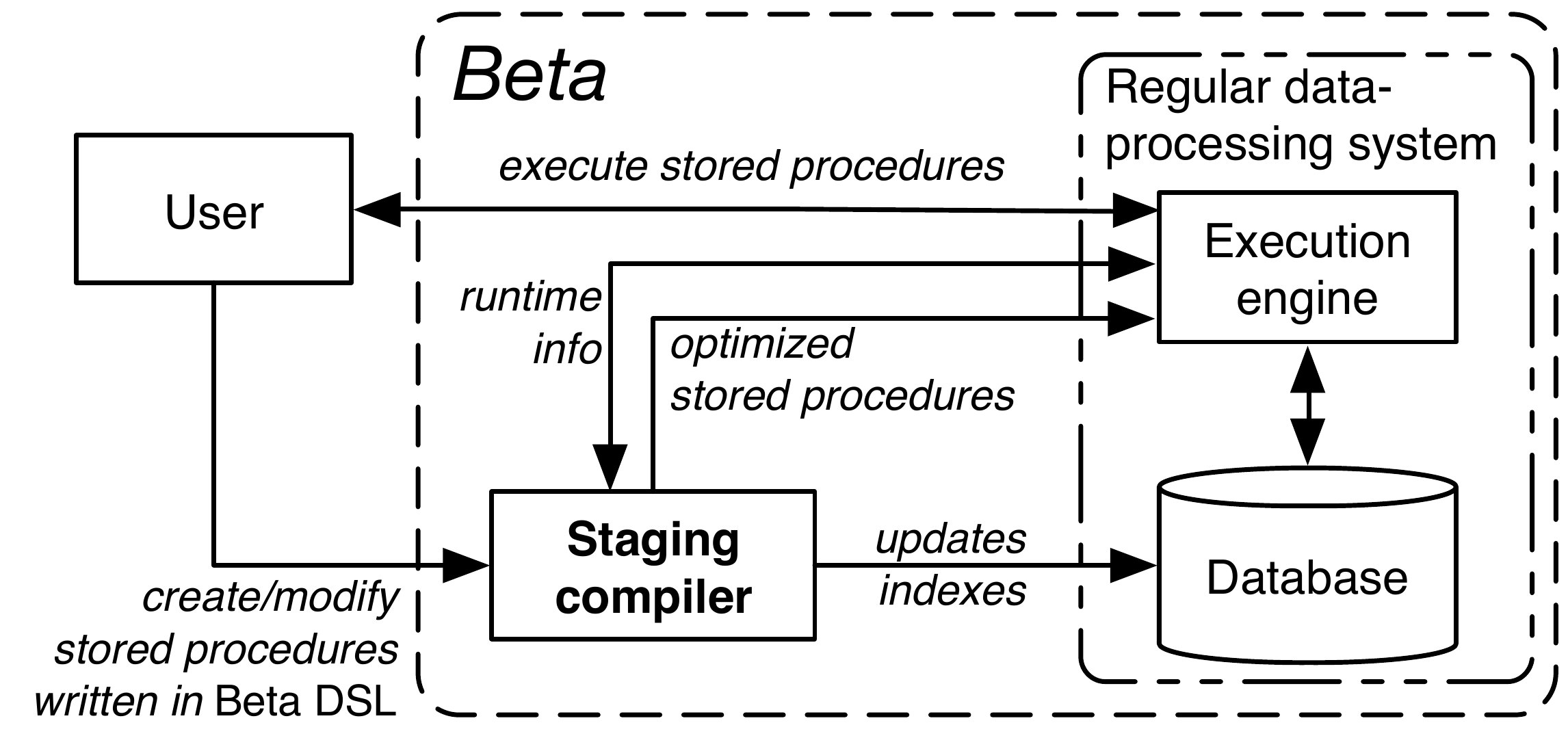}
\end{center}
\caption{The architecture of \systemName.}
\label{fig:arch}
\end{figure}

The proposed architecture for our domain-specific compilation suite is illustrated in Figure \ref{fig:arch}. We assume that users and applications interact with the database system via stored procedures representing the pieces of application programs: the database application code is moved into the server. This decision not only aligns with the assumption in many novel database systems \cite{hstore_rewrite} and minimizes the round-trips between user programs and the database system, but also allows for the necessary code optimizations and the dynamic loading of the freshly compiled code during the operation of the system. In addition, in \systemName, with the assumption of having the whole set of stored procedures that are known apriori, the whole application code is compiled together with generic database server code. When these two pieces are put together, the indirections are removed and more optimization opportunities are uncovered.

\systemName makes use of an extensible domain-specific language (DSL), called \systemName DSL, for writing database application programs. In any domain-specific compiler, the language in which the user programs are written defines the highest level of abstraction given to the compiler. It starts applying its optimizations from that high-level domain. It is therefore essential to have a limited language, but powerful enough to express the application programs. More details of \systemName DSL is described in section \ref{sec:dsl}. 

\begin{figure}[t!]
\begin{center}
\leavevmode
\includegraphics[width=\columnwidth]{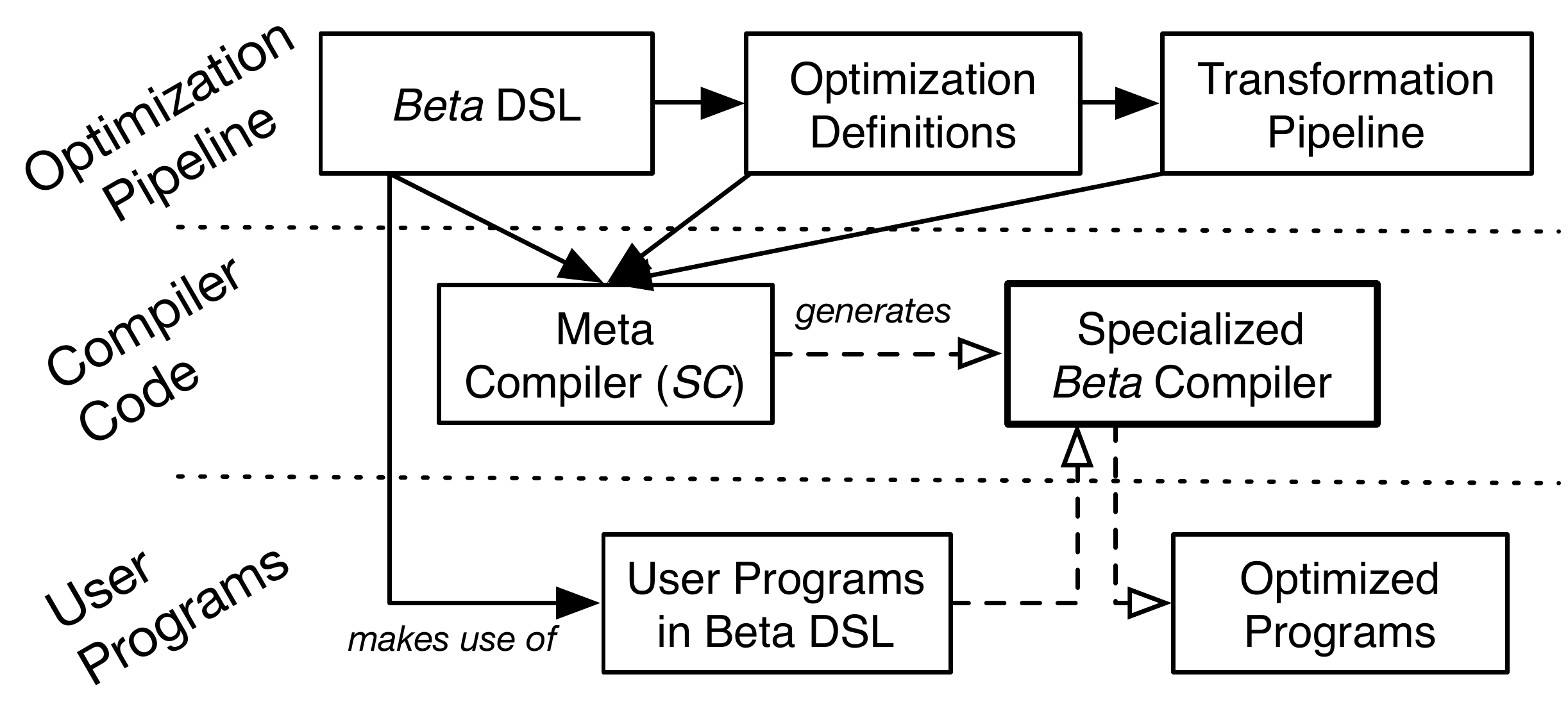}
\end{center}
\caption{The components of \systemName's staging compiler.}
\label{fig:scstages}
\end{figure}

\begin{figure}[t!]
\begin{center}
\leavevmode
\includegraphics[width=\columnwidth]{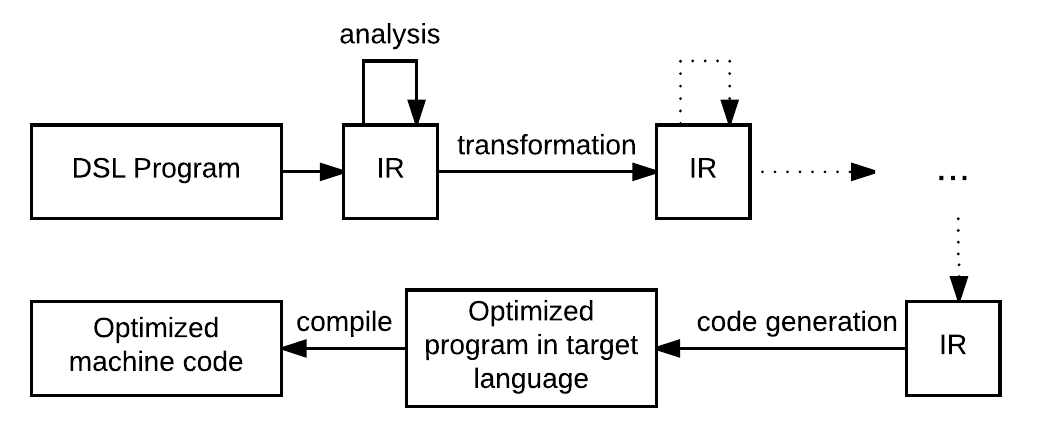}
\end{center}
\vspace{-4mm}
\caption{The compilation pipeline for a \systemName DSL program.}
\label{fig:pipeline}
\end{figure}

\systemName uses \SC \cite{sc}, a publicly available extensible meta-compiler and staging framework for Scala \cite{scala_overview}, to compile the programs written in \systemName DSL. \SC contains an extensible library for applying domain-specific optimizations and code generation for Scala and C++. Figure~\ref{fig:scstages} illustrates how \SC is used within  \systemName. Before database application programs can be written, the \systemName DSL must first be defined and embedded within the host language, Scala. The next step is to define the domain-specific optimizations and code transformation passes that are applied to this DSL. Given all this information, \SC generates the templates for intermediate representation (IR) \cite{duboscq2013graal, lms_opt_datastruct} nodes that will later be used to represent a DSL program during its compilation. The generated IR allows interleaving general purpose constructs (e.g., conditionals and loops) with domain-specific operators (e.g., joins and projections). \systemName automatically converts every optimization that is defined on the \systemName DSL into functions that manipulate IR nodes representing the program.
These optimizations are chained together in the form of a compilation pipeline as shown in Figure \ref{fig:pipeline}.
    
To maximize performance, the compiler outputs highly optimized low-level code and all high-level operators are replaced. Accordingly, within the same compiler IR we combine \emph{lowering} and \emph{optimization} passes \cite{lms_opt_datastruct, sc}. Lowering passes convert high-level nodes into their equivalent representation in a more concrete and lower level IR counterpart, while optimizations improve its performance without changing the current level of abstraction of IR graph. Before every lowering, it is necessary to perform all optimization passes in order not to miss any optimization opportunities. For example, join reordering and select-predicate pushdown are applied before lowering to the appropriate implementation. Optimization passes followed by the lowering passes are applied until the low-level code is generated.

The goal of \systemName is to produce highly optimized machine code. To achieve that, complex platform-specific optimizations must be applied for which domain-specific knowledge is useless. The optimizations, such as register allocation, instruction
scheduling, cache optimizations are already part of  general purpose compilers and  \systemName leaves these optimizations to a general purpose compiler. Instead, \systemName generates code in target languages such as C++ or Scala and gives it to the respective compiler so as to produce the optimized binary. This simplifies the design of \systemName while also making it platform agnostic.

Some optimizations that a human expert may suggest are impossible
for a compiler to find even in the best of all worlds, simply because they are
not {\em at all times}\/ correct and require some preconditions for their correctness.
As an example, the TPC-C code of \cite{hstore_rewrite} uses
fixed-size arrays to represent metadata tables (such as ``items'') that do not
change after a warmup phase of the benchmark.
This optimization contributes significantly to the reported performance improvements
over classical OLTP systems. But it
remains a cheat unless we find a way for the system to switch between a first metadata
loading phase and a second operational phase during which the metadata remains static and
only the other tables change. 
In \systemName, as shown in Figure~\ref{fig:arch}, we utilize information about the mutability of tables, configuration files, and metadata stored in the tables to produce the fastest possible code. This adds a second,
{\em  temporal}\/ dimension to {\em specialization by design}.

With the meta-compilation, even though the user program can be well modularized, the compiled program is more tightly coupled with the database system code. This is generally not a problem in \systemName as the programmers are not going to directly deal with the generated code and therefore, it does not affect the maintainability of the system. However, for any update to a component of the database system or a change in the application programs, we need to compile and deploy a new version. We need to make sure this is not on the critical execution path of the system, so as to avoid complete execution halts between updates.

After the first time that the database application goes into production, any further changes to the codebase go through \systemName to create the next specialized version of the application, which is a completely separate process from the running application. Then, the common techniques in \textit{continuous delivery} \cite{contdelivery} are used to push the updated application into production, assuming the old and new versions of the application access data from a region of memory shared by both. Any change to the underlying system or application program can have two implications: changing the application code and/or underlying data structures and indexes. If it is only the code, then only the new version of the code is dynamically loaded into the execution engine and the new requests are directed to this new version. When the underlying data structures and indexes change, a transition plan is required to move the data into the new version. 

The new indexes are created during the application update; after this point, only the instances of new application programs are started; transition then happens gradually until all instances of the old application programs have finished; finally, unused indexes are pruned. If creating a new index is required, a low-overhead snapshotting mechanism is used to capture the current data in the table as well as its subsequent changes during the index creation process, as requests are continuously handled by the application in production. Deleting an index is postponed until all the requests to the previous version of the application are processed. In a distributed setting, an application update could involve re-partitioning the data; this requires replication consistency during the update.

In the following sections, we go into more details of some of these components.
\section{A DSL for Database Applications} \label{sec:dsl}

A DSL is a language specialized for a certain domain. DSLs are often high-level and declarative.
A declarative interface exposes best the real intent of the programmer by hiding unnecessary details, e.g., loop counters.
The key to success for DSLs is to impose a principled way of structuring a program over which the compiler can reason. Recent advances in compiler technology \cite{meta_ocaml,compile_edsl, haskell_nest_para} and DSLs \cite{lms_hp1,lms_opt_datastruct,lms_original,sc,optiml} are leading to the counter-intuitive insight that automatically generated code can outperform expert handwritten code in some domains \cite{halide,spiral}.

\begin{figure}\begin{center}
\begin{lstlisting}[language=Scala]
trait Store[T] {
  // high-level operations
  def filter(f:T=>Bool): Store[T]
  def map[U](f:T=>U): Store[U]
  def fold[U](zero:U)(f:(U,T)=>U): U
  def join[U](s:Store[U],c:(T,U)=>Bool):Store[(T,U)]
  def groupBy[K](p:T=>K): Store[(K, Store[T])]
  def union[T](s:Store[T]): Store[T]
  // low-level operations
  def get[K](p:T=>K, key:K): T
  def slice[K](p:T=>K, key:K): Store[T]
  def range[K](p:T=>K, from:K, to:K,
             options:RangeOptions=DEFAULT): Store[T]
  def foreach(f:T=>Unit): Unit
  def insert(t:T): Unit
  def update[K](p:T=>K, key:K, updated:T): Unit
  def delete(t:T): Unit
}
\end{lstlisting}\end{center}\vspace{-6pt}
\caption{Interface for the \systemName DSL.}
\label{fig:dsl_interface}
\end{figure}

In \systemName, we focus on whole application programs; queries are just a part of them. Moreover, it uses a single language that unifies data manipulation and query operations. Having a unified language allows: \emph{i)} consistent syntax, \emph{ii)} unified compiler optimizations for queries and data manipulation programs. Our DSL (Figure~\ref{fig:dsl_interface}) is embedded in the Scala language \cite{scala_overview}, which offers an infrastructure for creating and optimizing DSLs \cite{sc}. The \systemName DSL has one dynamic data structure for tables (i.e., {\tt Store}) and contains two parts:\begin{itemize}
\item High-level functions that operate on relations and provide the same expressive power as SQL (with the Scala syntax). 
\item A low-level imperative DSL used for implementing the high-level query operators. 
\end{itemize}

The high-level operations can be expressed with the low-level DSL. Yet, distinguishing the two layers is important as some optimizations can only be applied at a higher level, but the programmer can choose to combine them. We now describe the DSL operations and how to {\em lower} the high level operations.

\paragraph{High-level DSL}
A high-level DSL gives more freedom to the compiler to choose the best implementation. SQL is one such a DSL for queries where the join implementation is left to the query optimizer.

Our high-level DSL is similar to Monad Calculus \cite{monad-calc-1} and collections in LINQ \cite{linq}.
Operation {\tt groupBy} assigns tuples to groups ({\tt p:T=>K}). Operation {\tt join} joins two relations and accepts a boolean function as join condition that, given a tuple from both relations, determines whether this pair is in the join result. The {\tt map} method is used for transformation (e.g., projection), {\tt filter} for selection, {\tt fold} for aggregation, and {\tt union} for combining two relations.

A query expressed in the high-level DSL is a sequence of function calls over {\tt Store} objects. This representation is convenient to rearrange operators (it is much harder in the low-level DSL). For example, we can apply classical query optimizations on this high-level DSL. These optimizations are typically done by a query optimizer and have been studied for more than forty years \cite{old_opt_relalg}.

\paragraph{Low-level DSL}
The low-level DSL contains fine-grained operations on the {\tt Store} objects. Selecting a single record in a unique index is done via {\tt get}, which accepts a function ({\tt p:T=>K}) to extract the search key from each record, as well as the target search key ({\tt key:K}). It should be guaranteed by design that the given key to {\tt get} is a unique key.
Operation {\tt slice} is similar to {\tt get} but used for cases where the given search key is not a unique key.
Operation {\tt range} applies range queries and the {\tt options} argument controls the inversion of the range and bounds inclusion.
Operation {\tt foreach} applies a function to each record of a {\tt Store}.
Finally, {\tt update}, {\tt delete}, and {\tt insert} modify the relation content.

The low-level DSL expresses query plans. Its interface forbids less efficient  functional style on {\tt Stores\footnote{Automatically transformed by deforestation, cf. \S\ref{sec:optimizations}.}}. The only supported operations are loops or operations on individual elements.
Although high-level data manipulation is functional (immutable data), the low-level DSL uses mutability for in-place updates with lazy re-indexing of the tuple in the {\tt Store}.
The low-level DSL still abstracts away indexes: the projections {\tt p:T=>K} define the potential partition columns, but do not impose an implementation. Indexes can encode complex operations that are further optimized for each phase specifically (see ~\S\ref{sec:auto_indexing}).
For example, in the TPC-C {\tt Delivery} transaction program, 
\begin{lstlisting}[language=SQL]
SELECT o_id FROM new_order WHERE d_id=:d
  AND w_id=:w ORDER BY o_id ASC LIMIT 1;
\end{lstlisting}
corresponds to the {\tt slice} operation followed by a minimum aggregator \footnote{The {\tt min} operation is implemented as an implicit conversion to an extended \systemName DSL that supports it, even though {\tt min} and other similar operations could be a part of the \systemName DSL and be implemented using the {\tt foreach} operation.}:
\begin{lstlisting}[language=Scala]
newOrder.slice(o=>(o.d_id,o.w_id),(d,w))
        .min(o=>o.o_id)
\end{lstlisting}
and the corresponding index could be implemented with a hash-table of heaps.

\paragraph{Transition from high-level to low-level DSL}
Once optimized, the high-level DSL is converted into low-level operators by the following rules:
\begin{itemize}
\item {\em Selection (filter)}: There are two cases for {\em filter}: if only one element is expected, we retrieve it with a {\tt get} operation, otherwise we iterate over all the elements matching the predicate using {\tt slice} for equality predicates, and {\tt range} to address all equalities and up to one inequality. If additional predicates exist, they can be applied to the output of {\tt slice} or {\tt range} operations. Exposing these  specific iteration operations helps generating specialized indexes (cf.~\S\ref{sec:datastructures}).
\item {\em Fold, projection (map), groupBy, union}: accumulate into a mutable variable or a regular Scala collection which then acts as a source.
\item {\em Join}: is converted into nested loops depending on the join predicate ({\tt foreach}, {\tt range}, {\tt slice}, and possibly {\tt get}).
\end{itemize}
Although the transformations are straightforward, combining them efficiently is not trivial.
As an example, consider the following SQL query (in the absence of NULL values) and its counterpart program:
\begin{lstlisting}[language=SQL]
SELECT SUM(p) FROM t WHERE c1=v1 AND c2<v2
\end{lstlisting}
\begin{lstlisting}[language=Scala]
t.filter( x => x.c1==v1 && x.c2<v2 )
    .map( x => x.p ).fold(0)((x,y) => x+y )
\end{lstlisting}
This program is lowered optimally using a range index; alternatively, we could use a hash index and test the rest of the filter predicate in the closure:
\\
\begin{lstlisting}[language=Scala]
// using range (tree-like)
var sum=0
t.range( x => (x.c1,x.c2), (v1,-INF),(v1,v2) )
       { x => sum+=x.p }
// using slice (hash-index)
var sum=0
t.slice(x => x.c1, v1,{ x => if (x.c2<v2) sum+=x.p })
\end{lstlisting}

Although high and low-level DSLs respectively bear resemblance with queries and query plans in the way they separate concerns, there exists a key difference that we can combine them and reuse the same constructs (control flow) across DSLs; the compiler can optimize across domains.
In the next section, we detail the optimizations that take place during the DSL conversion.

\section{Optimizations} \label{sec:optimizations}

We describe the key optimizations to achieve optimal transformation from a functional DSL into imperative programming (low-level DSL). This set of optimizations was obtained by reverse-engineering the TPC-C implementation of \cite{hstore_rewrite} and listed in Table \ref{tab:opts_used}.
We do not discuss common optimizations like dead code elimination (DCE) \cite{advancedcom}, common subexpression elimination (CSE) \cite{advancedcom} on both domain-specific operators and low-level constructs, e.g., removal of duplicate projection, selection, and joins, and loop-invariant code motion (computations are moved outside of loops when possible) \cite{advancedcom}. These optimizations are well studied and \SC \cite{sc} provides them out of the box.
Also, we do not cover high-level language optimizations; these are covered by classical query optimizers and are well studied elsewhere.

\subsection{Removing Intermediate Materializations} \label{sec:deforestation}

Composition allows us to write complex programs easily but it comes with a cost.
\begin{example} \label{ex:deforestation}
In a simple selection-projection query:
\begin{lstlisting}[language=Scala,mathescape]
customers.map(x => (x.credit-x.bal,
                    x.name.substring(0, 5))
        $\odot$.filter((credit, name) => credit < 47000)
        $\odot$.map((credit, name) => name)
        $\odot$.fold(0)((x, acc) => acc + 1)
\end{lstlisting}
each function iterates over the whole collection, and at each junction point ($\odot$) an intermediate collection is materialized (in memory).
\end{example}
In the above example, we map the original relation into a new collection. Then, we create a new subset of these elements, when we are only interested in the number of customers within a credit cap. Creating intermediate collections incurs large performance penalties, especially when the amount of computation per element is small.

\paragraph{Deforestation with staged data structures}
To remove materializations without changing the programming model, \systemName applies \emph{deforestation} (or \emph{fusion}) \cite{deforestation, jfppushpull} by using the techniques described in \cite{foldr-fusion-1}.
\begin{contexample}
If {\tt customer} is implemented as an array, the final program after {\tt deforestation} would become:
\begin{lstlisting}[language=Scala]
var res = 0 //accumulator of the fold
val len = customers.length
var i = 0
while (i < len) {
  val x = customers(i)
  // first map result
  val y = Entry(x.credit-x.bal, x.name.substring(0, 5))
  if (y._1 < 47000) { //filter
    val z = Entry(y._1) // second map result
    res += 1 //fold
  }
  i += 1
}
return res
\end{lstlisting}
\end{contexample}

Similar optimizations are applied to table joins, index operations, etc. For example, \systemName replaces relation joins by their appropriate join implementations.

\paragraph{Removing temporary records} 
Deforestation provides great performance improvements but cannot completely avoid materialization between function calls: an intermediate structure is constructed as the return value and immediately destructed by the next call.

\begin{contexample}
The value {\tt y} is created although the values of {\tt x} could be used directly. In the body of the loop we expect to obtain:
\begin{lstlisting}[language=Scala]
 while (i < len) {
   val x = customers(i)
   if (x.credit-x.bal < 47000) res += 1
   i += 1
 }
\end{lstlisting}
\end{contexample}

In order to remove intermediate materializations, one needs to be aware of the characteristics of the intermediate results (tuples). As creating intermediate results may have side-effects (e.g., a function passed to {\tt map} that also modifies a global variable in its body), this optimization requires data flow analysis and is not generally applicable.
Intermediate materializations can be eliminated in two ways:
\begin{itemize}
\item By using continuation passing style (CPS) \cite{appel_cps} where only relevant intermediate results are evaluated and used.
\item By breaking up structures (tuples) into individual variables, eliminating unused code, and fusing variables back together in merge-points of the control flow graph. This approach is taken by \systemName.
\end{itemize}

\subsection{Automatic Indexing} \label{sec:autoindex}\label{sec:auto_indexing}

In traditional databases, any SQL query could potentially be executed by a client. Therefore all tables could be mutated and automatic indexing is not directly possible without further assistance from the users, probably via user-specified annotations.
Choosing appropriate indexes is a difficult task that is left to a human operator, usually a database administrator, possibly guided by query optimizer hints and heuristics-based tools \cite{indexes_tool}.

\systemName gathers all the static information available in the database application and combines it with runtime information from the execution engine to guide the automatic indexing mechanisms, and automatically introduce indexes for the tables.
With \systemName, this is possible since, at any point in time, the set of possible programs is closed as they are defined by the current application. Furthermore, in \systemName DSL, all table accesses and writes are explicit and it is trivial to collect them. Given this closed world of programs and information about access patterns, \systemName infers the optimal indexes.

In the rest of this section, we talk about one such automatic indexing mechanism that is used by \systemName, but essentially the information gathered by \systemName can be fed into any other automatic index advisor tool \cite{indexes_tool}. Then, after proposing the proper indexes for an application, \systemName takes care of the necessary program transformations to employ these indexes, such as generating the proper hashing and comparison functions required by indexes.

\paragraph{Analyzing access patterns and data mutation}
The range of possible data structures for indexes is limited by the data mutations: if records are inserted or deleted, an immutable index (e.g., fixed-size array) can not be used; in the presence of updates, it depends on whether modified columns are part of the index.
To create efficient indexes, we are interested in the access patterns: partitioning ({\tt slice}), range scanning, minimum/maximum selection and uniqueness (primary or unique key).

In \systemName DSL, access patterns are detected based on the chain of Store operations (Figure~\ref{fig:dsl_interface}). For example, {\tt get} retrieves an element in a unique index and {\tt range} denotes a range selection.
The key insight is that indexes can be automatically generated during the compilation process. \systemName first searches for access patterns and mutations in the intermediate representation of all programs. Each relevant operation is registered with the associated table. After the analysis, every table in the phase is marked with mutability flags (insert/delete/update), and a list of candidate indexes (pattern and table columns). This information is then used to select indexes; heuristics are studied in the database literature~\cite{indexes_olap,indexes_olap2}.

\paragraph{Access index selection}
For in-memory databases, the clustering and sequentiality (reminiscent of disk-based DBMS) are less relevant as all locations can be accessed in constant time (without a seek penalty). Therefore, we can simplify primary and secondary indexes design by retaining only their uniqueness property.

To store and lookup data, \systemName relies on the most appropriate data structure: arrays, lists, heaps, B+trees, hash-sets, etc.
Each structure has different access time guarantees for lookup, iteration over a subset or all elements, and maintenance costs (in presence of update, insert and delete). 

The index creation strategy of \systemName is shown in Table~\ref{tab:indexDecision}; our insight is that we can nest well-known data structures into each other. For example, we obtain the minimum of a partition of the data using a heap nested in a hash-table. In a read-prevalent workload, this strategy always pays off because indexes are more often read than updated. A detailed example of applying the {\tt Automatic Indexing} on TPC-C benchmark is provided in section~\ref{sec:autoindexexample}.

\begin{table}
\vspace{4mm}
\caption{Decision-table for automatic index creation after program analysis in \systemName.} \label{tab:indexDecision}
\vspace{-2mm}
\begin{center}\small
\begin{tabular}{ll}\toprule
    \bf Store Operation & \bf Equivalent Index   \\ \midrule
    $<$dynamic table$>$ get & Unique hash            \\
    slice               & Non-unique hash        \\
    range               & Hash of B+ trees        \\
    min or max          & Hash of binary heaps   \\
    median            & Hash of two binary heaps (min/max)  \\
    $<$static table$>$ get  & Hash lookup Array                  \\
    (none)              & Linked list            \\ \bottomrule
\end{tabular}\end{center}
\vspace{-6mm}
\end{table}

\subsection{Automatic Indexing Example} \label{sec:autoindexexample}

\begin{figure}[ht]
\begin{center}
\leavevmode
\includegraphics[width=0.8\columnwidth]{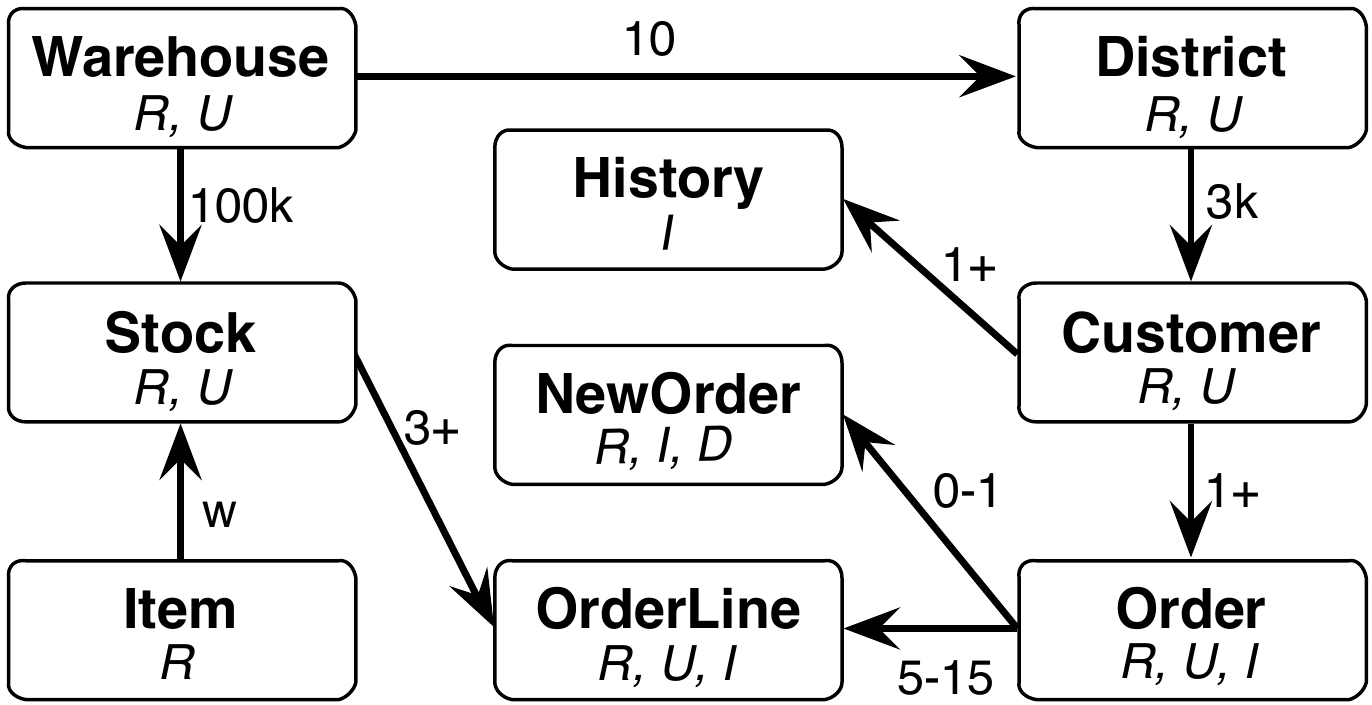}
\end{center}
\caption{Entity-relationship diagram for TPC-C benchmark and operations of the transactions on each table: R:read, U:update, I:insert and D:delete.}
\label{fig:tpccERD}
\smallskip
\smallskip
\end{figure}

We use the TPC-C transaction programs to illustrate the automatic index creation. 
The TPC-C benchmark consists of five transactions and the workload is a mix of them with specific ratios.
Two of them (OrderStatus and StockLevel) are read-only, others (NewOrder, Payment, and Delivery) also write into the database. The table schema of TPC-C is shown in Figure~\ref{fig:tpccERD}. 

The static analysis of TPC-C transactions shows that Warehouse, District, Item, Customer, and Stock tables do not grow or shrink in size. \systemName uses this information to convert the indexing mechanism for these tables to an array index over one (Warehouse, Item) or multiple columns (District, Customer, and Stock tables).

This conversion improves the performance: the array position is obtained with a numerical operation.
The ranges of indexed columns are also required; these are obtained by inspecting the data. \systemName maps the indexed columns into array position as follows:
\begin{itemize}
\item If the range (product) is much larger than the number of entries, we use a hash-table
\item If a single column is indexed, its values must be unique
\item If the index uses multiple columns, their combination must be unique, and therefore, \systemName constructs a mapping from the cross-product of the ranges to an integer value.
\end{itemize}

 For the rest of the tables, most of the queries are either lookups on the primary key or contain equality conditions. Hashing is, therefore, the preferred indexing method. The index columns are determined by the key of {\tt get} and {\tt slice} operations.
For queries containing inequality conditions, \systemName chooses between iterating the whole map when selectivity is low and using a B+ tree index when it is high. Additionally, a B+ tree index (possibly nested inside a hash-index) addresses all equalities and one inequality condition, and other conditions are verified on the returned tuples. Similar rules apply for minimum, maximum and median queries. \systemName uses a hash-table of min or max-heap indexes for faster lookup of minimum or maximum element in a partition, respectively. Moreover, a hash-table of two combined balanced heaps (one min-heap and one max-heap, where their sizes differ by at most one) is used for efficient lookup and maintenance of the median element in a partition.

\begin{paperexample} \label{ex:delivery}
Consider a simplified excerpt of the Delivery transaction for TPC-C benchmark in \systemName DSL:
\begin{lstlisting}[language=Scala]
for (d_id <- 1 to DIST_PER_WAREHOUSE) {
  val top_order:NewOrderEntry = 
     new_order.slice(x=>(x.d_id,x.w_id), (d_id,w_id))
              .min(x=>x.no_o_id)
  orderIDs(d_id - 1) = top_order.o_id
  new_order.delete(top_order)
}
\end{lstlisting}

The generated code contains two parts: the specialized data structure for {\tt NewOrder} tuples (NewOrderEntry) and the executable program code. The {\tt NewOrder} table is provided with index-specific methods for comparison and hash code. The generated code with reusable keys and automatic indexing ({\em without inlining}) is:

\begin{lstlisting}[language=Scala]
class NewOrderEntry(var o_id:Int=0, var d_id:Int=0,
                    var w_id:Int=0) extends Entry
type E = NewOrderEntry
object NO_Idx extends IndexFunc[E] { // slicing
  def hash(e:E) = do_hash(e.d_id,e.w_id)
  def cmp(e1:E,e2:E) = if (e1.d_id==e2.d_id &&
                           e1.w_id==e2.w_id) 0 else 1
}
object NO_Ord extends IndexFunc[E] { // ordering
  def hash(e:E) = 0 // unused
  def cmp(e1:E,e2:E) = signum(e1.o_id-e2.o_id)
}
val new_order = new Store[E]()
// index(1)=SliceMin: slice=NO_Idx, cmp=NO_Ord
new_order.index(NO_Idx, SliceHeapMinIdx, NO_Ord)
val param = new NewOrderEntry //parameter holder

// executable program: gathers the oldest NO for
// each district and removes it from new_order table
var x1: Int = 1
do {
  param.d_id=x1; param.w_id=w_id;
  //get the minimum NO using the first index
  val x2 =  new_order.get(1, param)
  val x3 = x2._1
  val x4 = x1 - 1
  orderIDs(x4) = x3
  new_order.delete(x2)
  x1 += 1
} while (x1 <= 10)
\end{lstlisting}

A hash-index is created over $(w\_id,d\_id)$; each bucket is a binary-heap where the order is given by the $o\_id$ column. Other optimizations in the above code are: reusable keys ({\tt param}) and record structure specialization ({\tt NewOrderEntry} class is generated). 
\end{paperexample}

In practice, our greedy strategy applied to 22 TPC-H queries created only 255 indexes for 176 relations involved in these queries (a single index for 107 relations, two indexes for 59 relations, and three indexes for 10 relations).
Collecting information about relative frequencies of index operations could help in other scenarios.

\subsection{Hoisting} \label{sec:hoisting}

In many programs, there are temporary objects that either have a complex internal implementation that is not worth inlining or require using a considerable amount of initialization or dynamically allocated memory. In these scenarios, \systemName does not see any benefit in inlining and co-optimizing these helper objects with the user-code. However, it re-uses these objects across multiple program executions by moving the helper object into the global scope if possible. Examples of such helper objects that can be hoisted include the temporary objects used for storing the search key fields to be passed to {\tt get} or {\tt slice} operations.

\subsection{Partial Evaluation} \label{sec:partialEval}

Partial evaluation \cite{jones1993partial} is a technique to achieve code specialization by using partially known program inputs to generate a new version of the program specialized for those inputs. Using the statically known input (e.g., configuration files), \systemName partially evaluates an input program and converts it to a specialized program.

For example, in regular expression matching, the pattern is usually a constant string. The regular expression matcher can be partially evaluated and, when combined with {\tt Hoisting}, moved out to the global scope. String formatter is another commonly used example where the format is normally a constant string. The string formatter can be partially evaluated and replaced with specific low-level string operations for that particular format. Then, it is called efficiently many times during the application lifetime, without the necessity to parse and execute the pattern during the execution. Consider the following line of code from the {\tt Payment} transaction of the TPC-C benchmark.
\begin{lstlisting}[language=Scala]
val h_data="%.10s    %.10s".format(w_name, d_name)
\end{lstlisting}
After {\tt Partial Evaluation}, the low level C code generated by \systemName is the following:
\begin{lstlisting}[language=Scala]
char h_data[24];
strncpy(h_data, w_name, 10);
strcat(h_data,"    ");
strncat(h_data, d_name, 10);
\end{lstlisting}

\subsection{Record Structure Specialization} \label{sec:recspec}
In \systemName, by default, data is stored in the form of general purpose records implemented as a hash table that maps column numbers to untyped data values. The records have to be general purpose, as the database contains several tables, each of which may contain a different number of columns of different types. However, after enabling  {\tt Record Structure Specialization}, \systemName generates specialized data structures for records of each table in the database. These specialized records contain the exact number of columns of the appropriate types for the table. Each column can be accessed directly, and as the correct type, without requiring a hash table. Moreover, \systemName rearranges the memory layout of the records to group the key fields together. This optimization results in better data locality and more efficient memory access as one level of indirection is removed.

\systemName first infers the schema of the table from the data loading operations. The schema comprises the list of types of individual columns of the corresponding table. Once the schemas of the tables are known, \systemName invokes a type inference logic to associate each instance of records in the application program with the correct schema. It is based on the table whose {\tt Store} instance produced/used the record instance in one of its methods. Finally, the generic record instances are replaced by the specialized ones, along with all associated methods (i.e., getters and setters). In addition, \systemName automatically generates the code for defining these specialized data structures, along with equivalent hashing, equality and comparison methods.

\begin{figure}[t]
\begin{center}
\leavevmode
\includegraphics[width=\columnwidth]{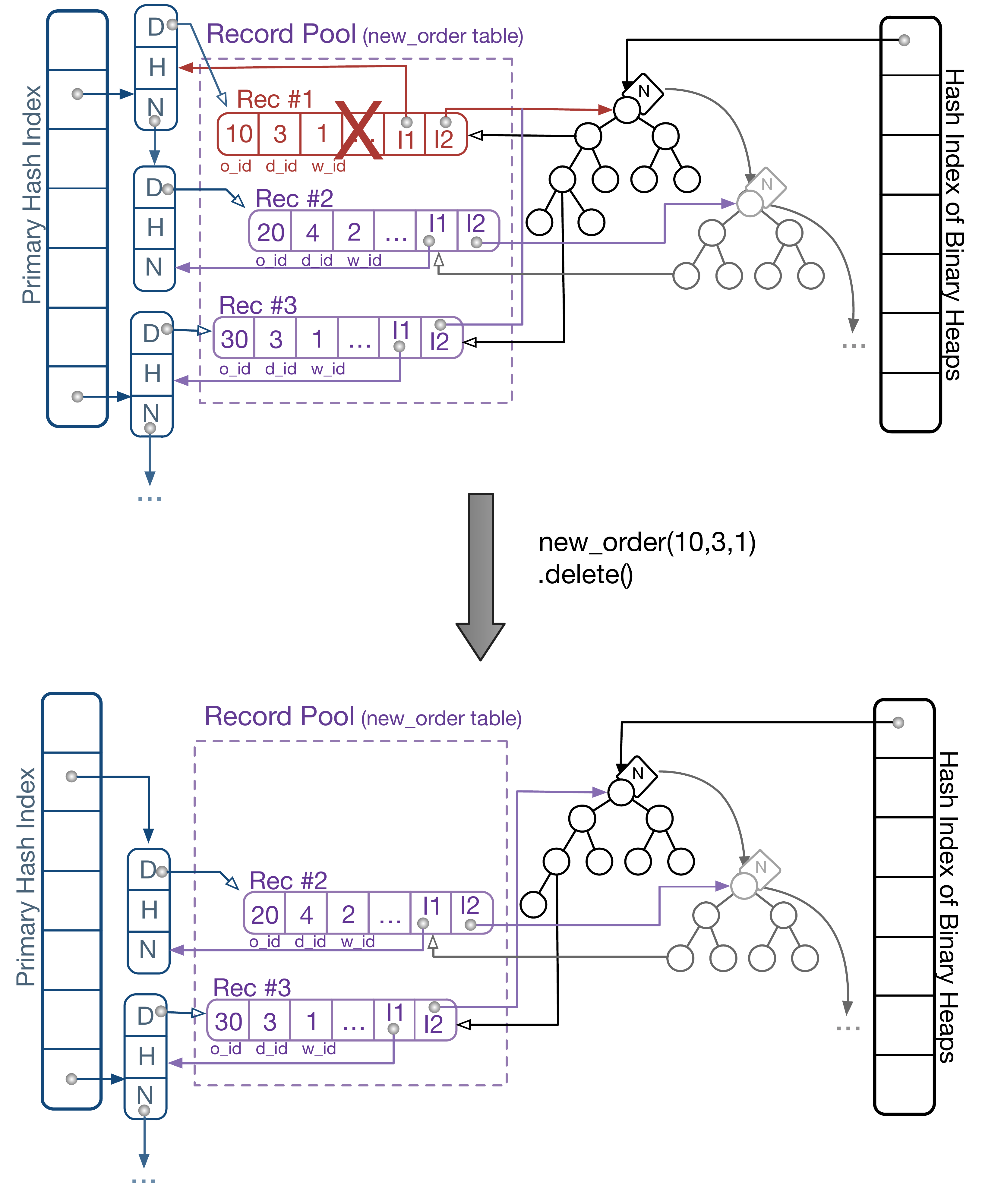}
\end{center}
\vspace{-4mm}
\caption{A sample schematic of multi-indexed data structures (IODS) generated by \systemName.}
\vspace{2mm}
\label{fig:datastructure_schematic}
\end{figure}

\subsection{Mutable Records} \label{sec:mutableRecs}
Typical database engines usually return a copy of the values when queried by a user for a multitude of reasons \cite{plprag}. Any changes to these values are not reflected back to the database unless followed by a subsequent update operation.
Consider the following code snippet from the {\tt Delivery} transaction in TPC-C:

\begin{lstlisting}[language=Scala]
val order=orderTbl.get(t=>(t.no_o_id,t.d_id,t.w_id), 
                          (no_o_id, d_id, w_id))
order.o_carrier_id = o_carrier_id
orderTbl.update(t => (t.no_o_id, t.d_id, t.w_id),
                     (no_o_id, d_id, w_id), order)
\end{lstlisting}

In the above code, an {\tt Order} is looked up, gets updated and is stored back in the table. There are several optimization opportunities: 1) the first line returns a copy of the {\tt Order} record, 2) the third line looks up the {\tt Order} record again to update it, and 3) the update does not affect any index as the updated field is not indexed. Compiled by \systemName, the equivalent compiled code is the following:

\begin{lstlisting}[language=Scala]
val orderRef = orderTbl.getRef(t => (t.no_o_id,
              t.d_id, t.w_id), (no_o_id, d_id, w_id))
orderRef.o_carrier_id = o_carrier_id
\end{lstlisting}

In this optimized code snippet, all the previously mentioned optimization opportunities are used. The first two opportunities are discussed here, and the last one in the next section.

In short, \systemName avoids copying values and subsequent lookups by returning a reference to the actual indexed data instead of a copy when it is safe to do so. In this example, after analyzing the input code, \systemName finds out that the {\tt Order} record is used only for updating a non-key field, so it uses {\tt getRef} instead of {\tt get} to directly get a reference to the internal {\tt Order} record. Subsequently, the update on {\tt o\_carrier\_id}  gets applied directly to the {\tt Order} record. 
Generally, update operations still have to be invoked on the table to update any index meta-information; but this can be done now without another lookup owing to back-references (see \S\ref{sec:datastructures}) employed by \systemName. However, in this example, the {\tt update} operation is also completely eliminated, by another optimization explained next (\S\ref{sec:dead-index}).

\subsection{Removing Dead Index Updates} \label{sec:dead-index}

Each {\tt update} operation on a Store results in the propagation of the update to the underlying indexes created for it. However, doing all these updates is not always necessary as not all indexed columns are altered every time. For every update operation,  at compile time, \systemName tracks which columns are modified. As it knows what columns are indexed by each indexes, it replaces the update operation on the table with those on only the relevant indexes. Moreover, with the {\tt Mutable Records} optimization (\S\ref{sec:mutableRecs}), the primary index is not updated as well if the primary key columns are not updated, and the entire update operation on the table is avoided altogether as in the example above.

\section{Inside-Out Data Structures}\label{sec:datastructures}

The execution engine of \systemName requires a data structure that implements all the operations in the \systemName DSL (\S\ref{sec:dsl}) and supports all the optimizations described in section \ref{sec:optimizations}.
\systemName uses a multi-indexed data structure, called \IODS, that exposes its internals to the \systemName optimizations and does not follow the information hiding principle as in existing alternatives. The exposed internals help \systemName to use \IODS effectively. An \IODS is an implementation of the {\tt Store} in the \systemName DSL and is composed of a record pool and one or more indexes defined on top of it.  All the indexes defined on an IODS share the same records from the pool.  \systemName knows how these records are used by the application programs as well as by \IODS. In addition, these records are aware of their memory layout within \IODS at runtime.

The mechanism of sharing records among indexes taken in \IODS is different from the mainstream approach of separating keys and values. The benefit of this approach is that unnecessary data copies are avoided and there is a single location to apply all the updates to the data, even though the index meta-information still needs to be updated. It may appear that, by not separating the key for indexes, \systemName loses data locality for the hash and comparison functions on the key. However, this is not the case, as {\tt Record Structure Specialization} (\S\ref{sec:recspec}) takes care of putting the key fields near each other in the memory layout for the record structure.
Another disadvantage of the record pool approach is that indexes can no longer use generic hashing and compare functions, which are more convenient for the human programmer as the key fields vary between different types of records and indexes. On the other hand, if specialized hashing and compare functions are generated automatically, then the performance is improved by tuning these functions for their respective usages. This optimization is done nonetheless by \systemName and, therefore, this is not an additional overhead for the record pool approach.

\systemName takes advantage of the knowledge of how the underlying data structures are used in two ways. Firstly, it applies specialized data structure tuning by removing the unnecessary features of the data structure, or replace it with a better one. \IODS relies on \systemName to pick the most appropriate indexes for the application, which is explained in {\tt Automatic Indexing} (\S\ref{sec:autoindex}). Secondly, it takes over some parts of the implementation and bypasses the data structure as in the case of optimizations such as {\tt Mutable Records} optimization (\S\ref{sec:mutableRecs}).

By having records that contain meta-information regarding their position within the indexes, subsequent index operations on the same record are performed without having to look them up again. To realize this, the records in \systemName have back-references to their corresponding {\em index containers}, i.e., the closest sub-structure in the index that has a reference to the record. Moreover, each read operation from an index returns references to one or more of these complete records, instead of a copy, when instructed by \systemName. This works in conjunction with the knowledge of how a particular record is used within the application program.

As \systemName knows how each index handles individual records, it efficiently applies the {\tt Mutable} {\tt Records} (\S\ref{sec:mutableRecs}) and   {\tt Removing} {\tt dead index} {\tt-updates} (\S\ref{sec:dead-index}) optimizations. For an update operation, \systemName keeps track of modified columns at compilation time, and updates only the indexes referring to these columns. In addition, the {\tt update} and {\tt delete} operations become much cheaper to apply, as the indexes avoid any further lookup by using the back-references. For example, a delete operation in an index backed by a (doubly) linked-list becomes an O(1) operation instead of O(n) that would be required to look it up again. 

Figure~\ref{fig:datastructure_schematic} illustrates how table records are referenced by multiple indexes. 
This \IODS is generated for the {\tt NewOrder} table in TPC-C benchmark, and is automatically created based on the access methods on this table throughout the benchmark. Each cell is labeled with a letter, D, H, N, or I, which correspondingly stand for Data, Hash, Next pointer, and Index. The three fields in each {\tt NewOrder} are {\em order ID} ({\em o\_id}), {\em district ID} ({\em d\_id}) and {\em warehouse ID} ({\em w\_id}). The primary index is created on all of these fields and is chosen to be a hash-index. The secondary index is created on {\em d\_id} and {\em w\_id} and is a hash-index of binary heaps, where each binary heap maintains the minimum {\tt NewOrder} record ordered by {\em o\_id}. Each {\tt NewOrder} record stores back-references to its corresponding index containers. 

Figure~\ref{fig:datastructure_schematic} also shows how a {\tt delete} operation on a {\tt NewOrder} record benefits from the available back references to the index containers for applying the operation without a prior lookup into the indexes. In this example, a  {\tt NewOrder} record is first looked up from its secondary index, and with the {\tt Mutable Records} optimization enabled, a reference to the actual record is returned. It is this reference that is passed to the {\tt delete} operation on the \IODS, which in turn passes it to the {\tt delete} operation on the indexes. Since each index receives the actual reference to the entry to be deleted, it can use the back-references to find its container for that record. For the primary hash-index, the container forms a part of the doubly linked collision chain for that hash bucket, and removing the container is straightforward as in any doubly linked list. For the secondary index, the container is part of a heap that resides in a hash bucket. Its deletion requires a lookup in neither the hash-index nor the heap, and is a standard remove operation on the heap.

\section{Evaluation}\label{sec:evaluation}

To evaluate \systemName, we use two benchmarks: TPC-C and trigger programs generated by DBToaster \cite{dbt_delta} (r2827) for the incremental view maintenance of TPC-H queries.
All experiments are performed on a single-socket Intel\textsuperscript{\circledR}  Xeon\textsuperscript{\circledR} E5-2620 (6 physical cores), 128GB of RAM, Ubuntu 16.04.3 LTS, GCC 4.8.4, and Oracle Java(TM) SE Runtime Environment (build 1.8.0\_111-b14).
Hyper-threading, turbo-boost, and frequency scaling are disabled to achieve more stable results.

We used \SC 0.1.4, running with Scala 2.11.2 on the Java Hotspot 64-bit server VM having 64 GB of heap memory, and the generated code was compiled with the \mbox{{\tt -optimise}} option of the Scala compiler and \mbox{{\tt -O3}} for GCC. We run all benchmarks 5 times and report the median of last 3 measurements.
Variance in all experiments is less than 3\% so we omit it from the graphs.

\subsection{Performance Model}
To evaluate \systemName, we study the impact of individual optimizations by enabling different combinations of optimizations on the same application. However, to compare the impact of individual optimizations, we need a base case. Naturally, the first thing that comes to mind is to compare the scenario with only one optimization against the unoptimized application, for each optimization. We consider the ratio of throughput of the application after applying the optimization to that of the original application, as the impact of that particular optimization. To be precise, if $t_0$ is the total time for executing the unoptimized application, $t_X$ is that after applying optimization $X$, and $s_i$ is the speedup provided by $X$ through optimizing a sub-program (block)  $b_i$ that constitutes a fraction $f_i$ of the total time $t_0$ of the unoptimized application, then
$t_X := t_0 * \sum {\frac{f_i}{s_i}}$.

The impact of optimization $X$ is given by $\frac{t_0}{t_X}$, as the ratio of throughputs is the inverse of that of execution times. We  infer that 1) keeping the fraction of blocks $f_i$ constant, if the speedup $s_i$ is increased, the overall impact of the optimization also increases, and 2) increasing $f_i$ for blocks with large speedup $s_i$ also increases the overall impact due to the restriction that $\sum{f_i}$ is one. In other words, the higher the speedup $s_i$ and the fraction $f_i$ of blocks with high speedup, the higher would be the impact of the optimization.

However, this model does not do justice to some optimizations that are crucial for the most optimized code but are insignificant in the original application. This happens because the impact of an optimization depends not just on what it can do ($s_i$), but also on where it can be applied ($f_i$). For example, a block of code $b_k$ can have a small fraction $f_k$ in the unoptimized code, but as other optimizations are applied, $f_k$ can increase significantly if these optimizations improve other blocks and leave $b_k$ relatively untouched. At this point, the impact of applying some optimization $Y$ that improves $b_k$ would be much higher compared to when it was applied in the original application. This phenomenon is orthogonal to that where two or more optimizations have an enabling or a disabling effect on each other because a particular block can be optimized by all of them. In view of this, we chose the application obtained after applying all the optimizations to be the base case. To study the impact of an optimization $X$, we disable it and compare the throughput with that of the most optimized case.

\begin{figure}[!t]
\begin{center}
\leavevmode
\includegraphics[width=\columnwidth]{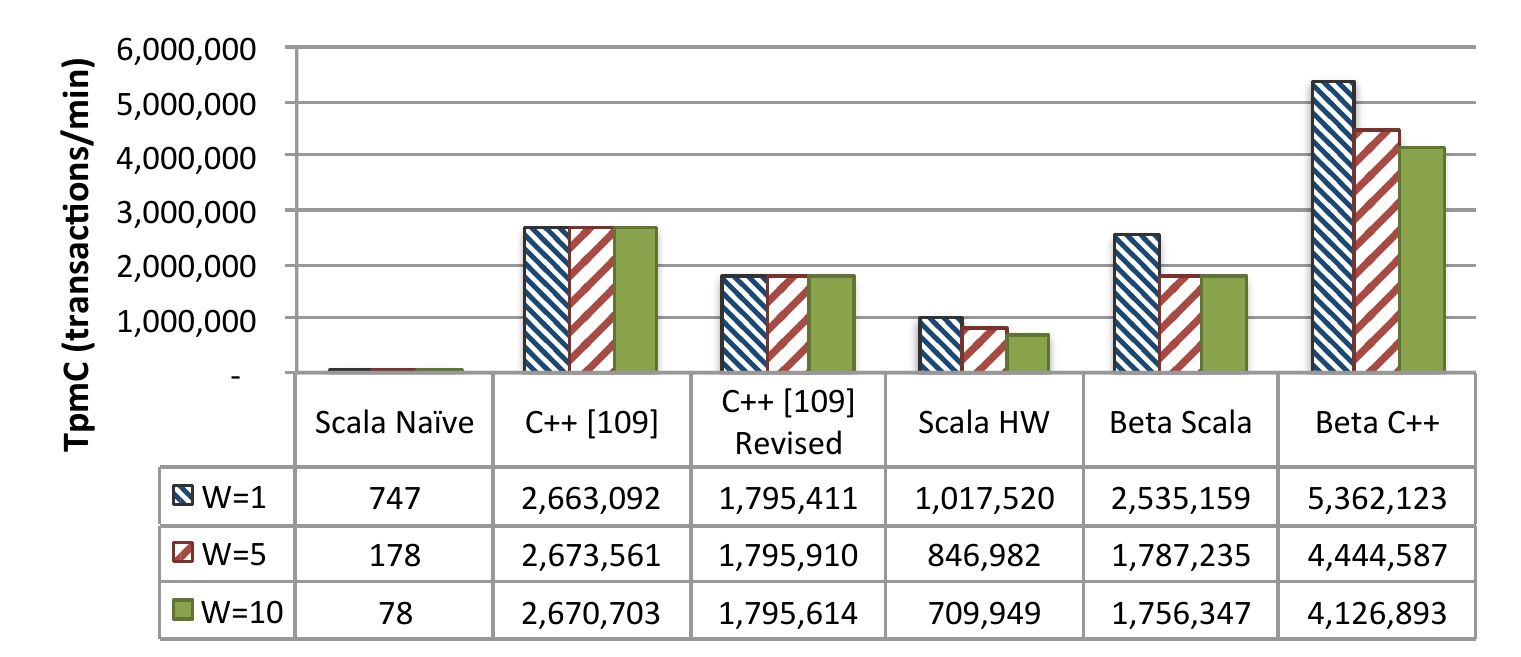}
\end{center}
\caption{TPC-C benchmark results. Comparison is made based on TpmC for different scale-factors (W). TpmC is the primary metric for TPC-C benchmark, which denotes the number of new-order transactions per minute.}
\label{fig:tpccBench1}
\end{figure}

\subsection{TPC-C Benchmark} \label{sec:evaluation_tpcc}
We implemented TPC-C transaction programs by a syntactic rewrite of SQL into \systemName DSL. The conversion is not yet automated but the correspondence is straightforward (see \S\ref{sec:dsl}).
The TPC-C benchmark specifies the meta-data characteristics and the input parameters for each transaction. This includes the relative size of each table and the expected number of records returned by a query. An expert programmer can, therefore, overuse this information to build a fixed-size implementation that is not otherwise correct. Conversely, we use \systemName to infer this information automatically and adapt the compiled program to the underlying data via adding annotations to the program. The results presented in this section are for the single-threaded implementation; those for the multi-core experiments are presented in section~\ref{sec:concurrency}.
We compare the code generated by the \systemName compiler with the expertly-written implementation of \cite{hstore_rewrite} and other implementations mentioned below. All benchmarks are restricted to single-site execution. The benchmark is run with eight million transactions.

\begin{figure}[!b]
\begin{center}
\leavevmode
    \includegraphics[width=\columnwidth]{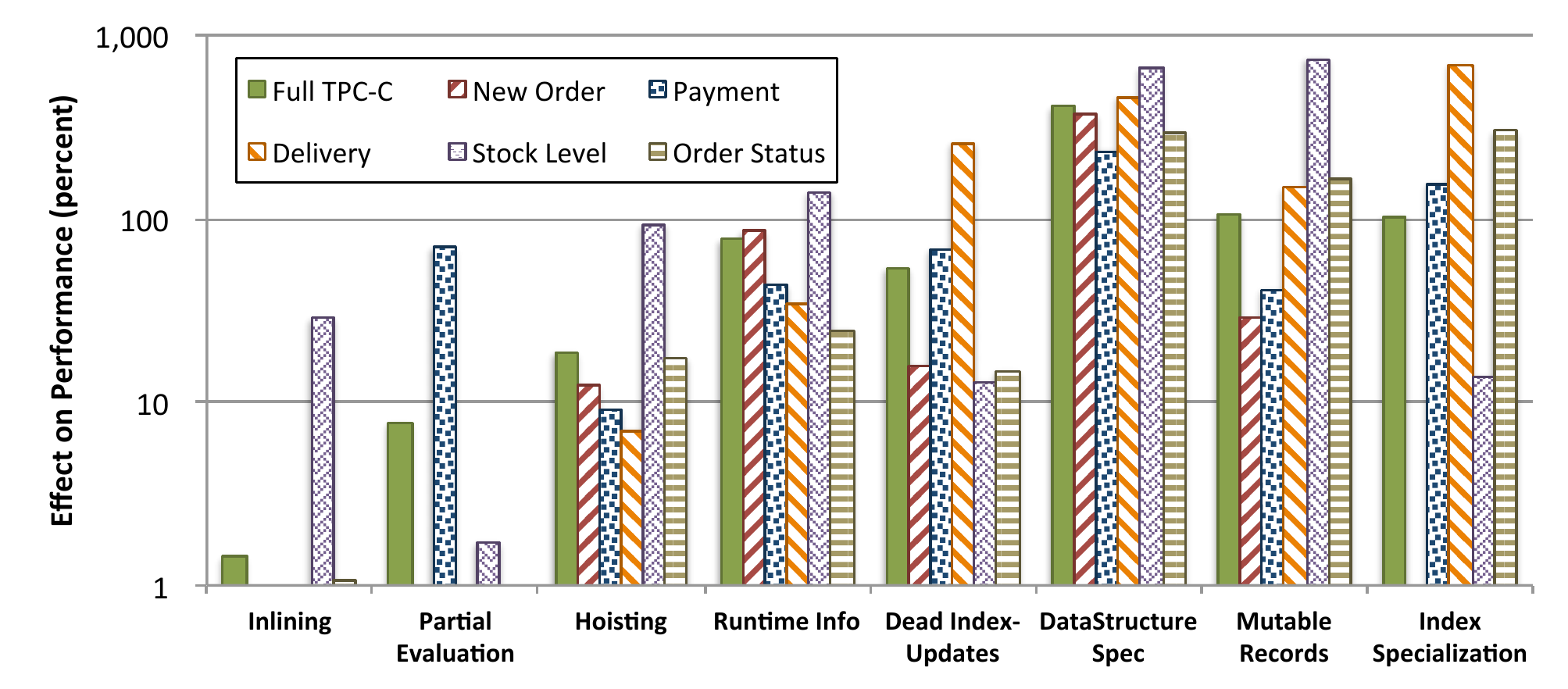}\vspace{-6pt}
\end{center}
\caption{The impact of important optimizations applied by \systemName to TPC-C.
}
\label{fig:tpccOptImpacts}
\end{figure}

\begin{itemize}
\item {\em Scala na\"{i}ve}: a direct conversion from SQL to Scala using only the Scala collections without any indexes on the tables.
\item {\em C++ \cite{hstore_rewrite}}: the C++ hand-optimized code of \cite{hstore_rewrite}. As mentioned in \cite{hstore_rewrite}, slight changes to the TPC-C benchmark were necessary; we applied the same change-set to other queries to have a fair comparison.
\item {\em C++  \cite{hstore_rewrite} Revised}: we corrected all the bugs of \cite{hstore_rewrite}:
converted static structures to allow experiments longer than one second, 
corrected the faulty implementation of B+Tree,
enabled a costly and necessary condition branch (Payment),
replaced a dummy value by an actual minimum lookup (Delivery), and 
added a count-distinct (Stock-level).
\item {\em Scala Hand-written}: the counterpart implementation of C++ Revised in Scala with hand-optimized code and manually tuned indexes.
\item {\em \systemName Scala/C++}: the program resulting from staged compilation. The \systemName input transaction code is the same as {\em Scala na\"{i}ve}, except it uses \systemName DSL instead of standard Scala collections. The compiled programs for both Scala and C++ output languages are measured.
\end{itemize}

We also compared it against other OLTP systems and found that they were one to two orders of magnitude slower. However, we do not present the results here as it is unfair to compare the code generated by \systemName to full OLTP systems whose main focus is not just obtaining the best performance possible, but also other objectives such as maintainability, support for ad-hoc transactions, etc. 

\paragraph{Results analysis}
Figure~\ref{fig:tpccBench1} shows that staged compilation of transaction programs produces C++ code that is twice as efficient as the hand-optimized C++ code written by experts. According to Table~\ref{tab:opts_used}, not only are some optimizations missed by the experts, but also \systemName applied the optimizations more thoroughly. In addition, Figure~\ref{fig:tpccBench1} shows that an optimized TPC-C implementation using \systemName in a high-level language (i.e., Scala) is not more than twice as inefficient as its C++ counterpart. Thus, \systemName optimizations are more relevant than its target language. The figure also shows a performance degradation with an increase in the number of warehouses. This happens due to the cache misses associated with processing data from different warehouses. However, the C++ implementation of \cite{hstore_rewrite} does not experience this as their B+Tree Indexes are more cache insensitive compared to our Hash Indexes. 

\begin{figure}[!b]
\begin{center}
\leavevmode
\includegraphics[width=\columnwidth]{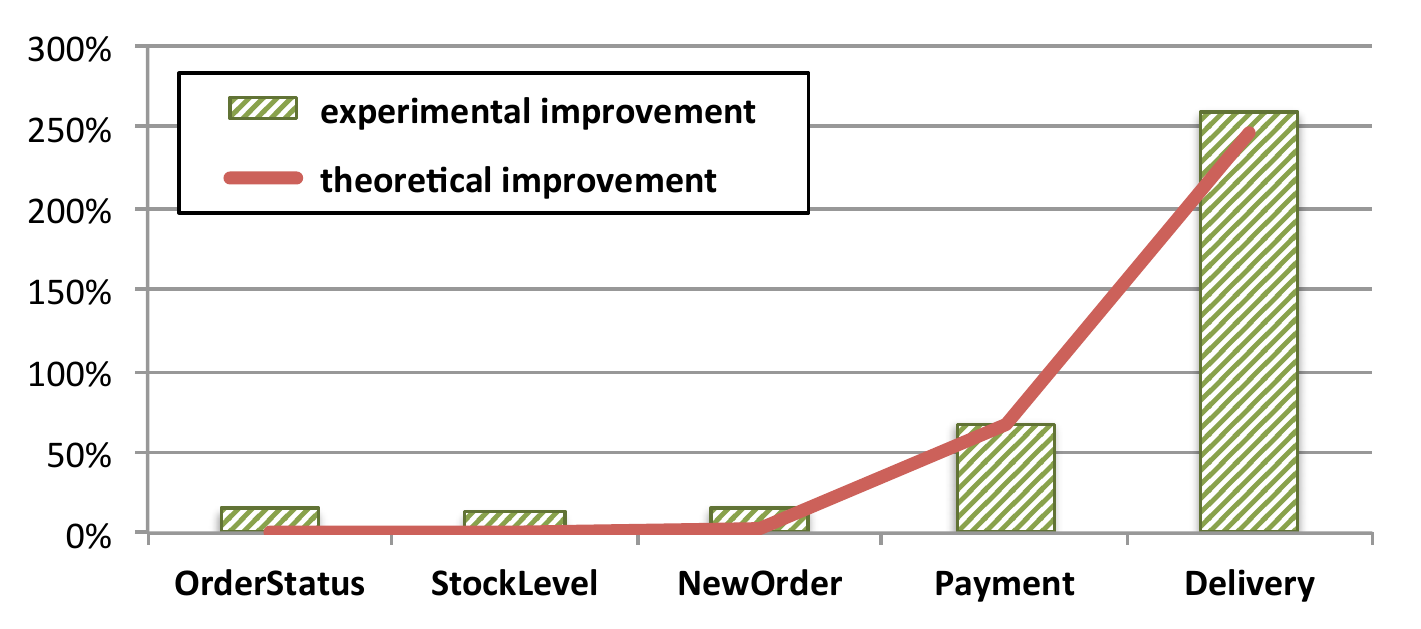}\vspace{-6pt}
\end{center}
\caption{Correlation between theoretical and experimental performance improvement by Dead Index Update on TPC-C transactions. The theoretical improvement is computed based on the fraction of updates in each transaction.}
\label{fig:tpccDeadIdx}
\end{figure}

The breakdown of the impact of individual optimizations for TPC-C is shown in Figure~\ref{fig:tpccOptImpacts}. In addition, we also show the breakdown for each of the individual transactions. Naturally, the impact of different optimizations differs across transactions according to their characteristics. First, let us consider the {\tt Removing} {\tt Dead} {\tt Index} {\tt Updates} optimization. In the TPC-C benchmark, {\tt Delivery}, {\tt NewOrder}, and {\tt Payment} transactions contain update operations to non-key columns. All of these updates are dead if the {\tt Mutable Records} optimization is enabled. Figure \ref{fig:tpccDeadIdx} confirms the correlation between the impact of {\tt Dead Index-Updates} and the fraction of time spent in update operations in each of these transactions in the absence of this optimization. If $f_u$ is the fraction of update operations, after plugging in $\infty$ for $s_u$, we get the theoretical impact of the optimization to be $(1-f_u)^{-1}$.

\begin{figure}[!t]
\begin{center}
\leavevmode
\includegraphics[width=.38\columnwidth]{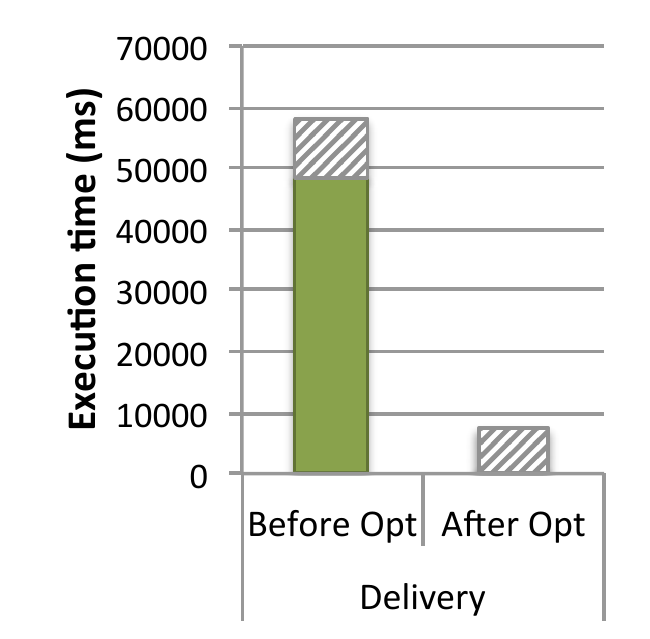}\hspace*{-0.9em}
\includegraphics[width=.32\columnwidth]{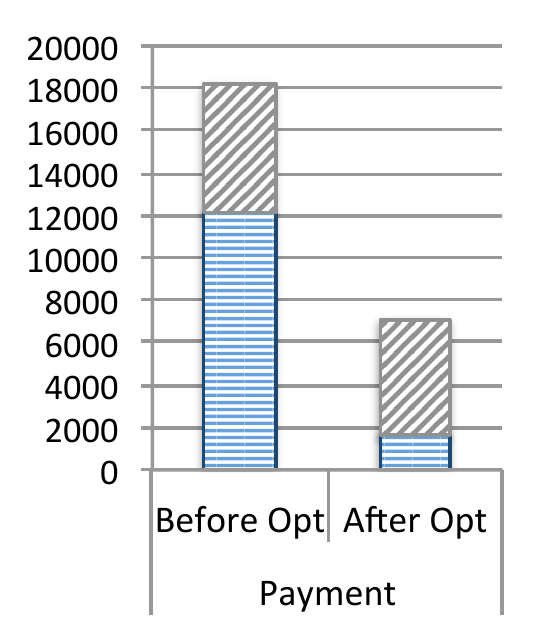}\hspace*{-0.9em}
\includegraphics[width=.32\columnwidth]{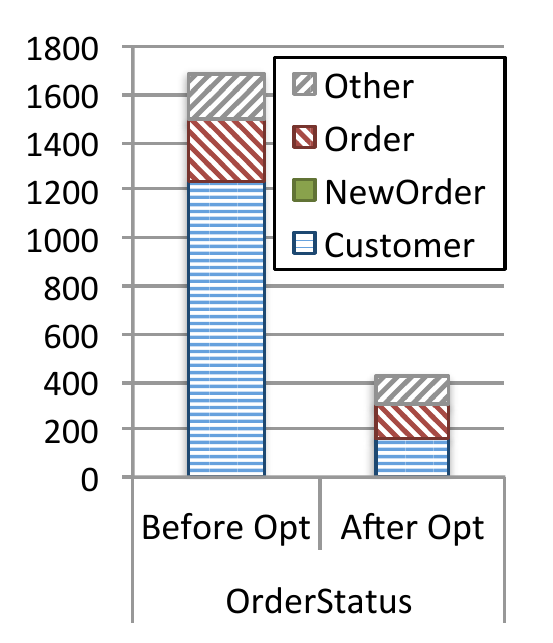}
\end{center}
\vspace{-2mm}
\caption{Breakdown of the amount of time spent on operations of each table for TPC-C transactions with minimum, maximum and median operations, before and after applying {\tt Index Specialization}.}
\label{fig:tpccSpl}
\vspace{2mm}
\end{figure}

Next, we have {\tt Index Specialization} (see \S\ref{sec:auto_indexing}) where \systemName introduces min-index for the {\tt NewOrder} table, max-index for the {\tt Order} table, and median-index for the {\tt customer} table.  Transactions {\tt Delivery}, {\tt OrderStatus} and {\tt Payment} use these indexes for efficiently performing their operations. Figure \ref{fig:tpccSpl} shows the composition of these transactions before and after applying this optimization. We can see that in {\tt Delivery}, the operations to find the oldest {\tt NewOrder} constitute roughly 89\% of the transaction. Introducing a min-index speeds up this block by a factor of 290, thus achieving an overall speed up of around 800\%. In {\tt Payment}, finding the median {\tt customer} takes roughly 67\%. With the median-index, it is 8x as fast, giving an overall improvement of 143\% to the transaction. {\tt OrderStatus} contains two blocks, one finding the median {\tt customer} by name (74\%) and the other finding the latest {\tt Order} (13\%). By introducing the median and max indexes, respectively for these blocks, we get a speedup factor of 8 and 1.5, respectively, contributing to an overall speedup of 250\%.

 Among all transactions, {\tt StockLevel} creates the maximum number of temporary objects with 200 {\tt stock} records to be used as temporary keys for primary key lookup. These {\tt stock} records comprise many string columns and are relatively more expensive to construct. Their construction and destruction form a significant portion of the {\tt StockLevel} transaction, and therefore, when {\tt Hoisting} moves these temporary objects out, this transaction has the highest impact. 

\begin{figure*}[t]
\begin{center}
\leavevmode
\includegraphics[width=\textwidth]{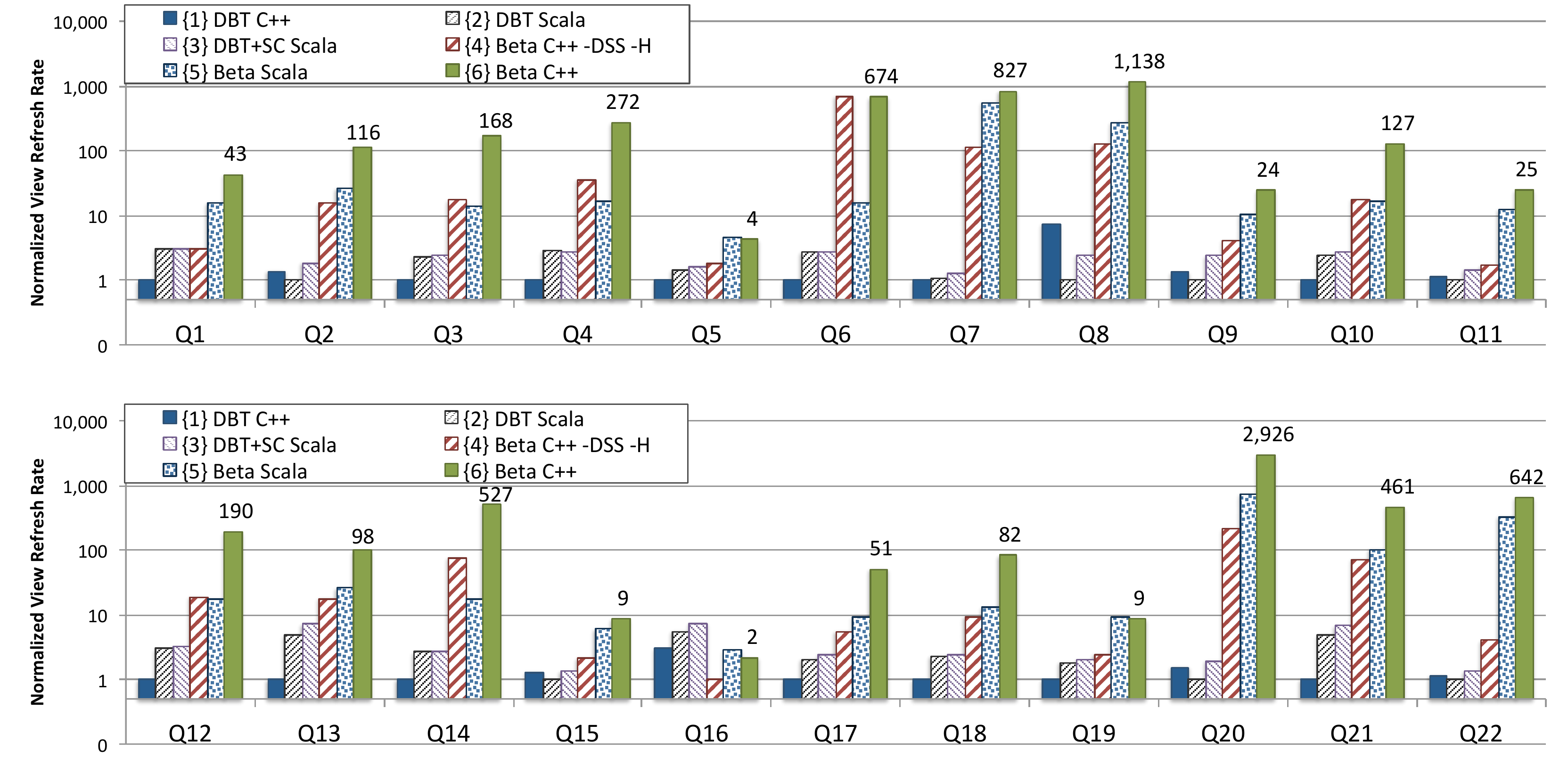}\vspace{-6pt}
\end{center}
\caption{Normalized view refresh rate for IVM triggers in TPC-H queries.} \label{fig:dbt1}
\end{figure*}

 Without {\tt Partial Evaluation}, about 50\% and 20\% of the {\tt Payment} transaction are taken up by the two string formatting functions ({\em sprintf}). However, this optimization evaluates the format string and converts these operations into efficient string manipulations, speeding them up by 3x and 2x respectively, giving an overall speedup of 75\% to the {\tt Payment} transaction. 
 
 With {\tt Runtime Information} (see \S\ref{sec:arch}), \systemName infers the approximate number of records in all tables, and pre-allocates memory for index structures. This avoids costly index resizing at runtime. Furthermore, it also infers that no records are inserted to or deleted from the {\tt warehouse}, {\tt district}, {\tt item} and {\tt customer} tables after the data loading phase. This enables \systemName to generate array indexes for these tables with efficient mapping from key columns to array positions that guarantee constant time lookup.

\subsection{View Maintenance Triggers for TPC-H} \label{sec:evaluation_dbt}

We see DBToaster \cite{dbt_delta,dbt_agile} as a generator for a class of trigger programs: the DBToaster front-end turns classical complex queries into triggers of multiple smaller-step materialized view updates.
We automatically rewrite such programs in \systemName DSL. Our main transformation consists of less than 70 lines of Scala.

We compare the \systemName DSL with the original DBToaster back-end (M3 to Scala).
Our compiler is shorter to write (Table~\ref{tab:loc_dbt_compiler}) and produces more efficient code.
To show its efficiency, we compare the original DBToaster back-end with our compiler.
In this benchmark, we measure the average throughput (during 1 minute) of {\em incremental view maintenance} (IVM) on a 200MB \mbox{TPC-H} dataset.
All the tables are initially empty, and the task is to compute the query result after receiving a new tuple for each of the base relations present in the query. The order of inserting the tuples is consistent among different experiments.
We do not compare with other systems as none implements similarly aggressive incrementalization techniques.
In Figure~\ref{fig:dbt1} we compare different compiler implementations for TPC-H trigger programs:\begin{itemize}
\item \{1\}{\em DBT C++}: original DBToaster C++ compiler.
\item \{2\}{\em DBT Scala}: original DBToaster Scala compiler.
\item \{3\}{\em DBT+\SC Scala}: we modified the low-level Scala code generated by DBToaster to use the generic optimizations of the \SC framework (DCE, CSE, and code motion).
\item \{4\}{\em \systemName C++ -DSS -H}: C++ program generated by \systemName after disabling {\tt Data Structure Specialization} and {\tt Hoisting}. 
\item \{5, 6\}{\em \systemName Scala/C++}: the program resulting from staged compilation. The input code for \systemName is similar to {\em DBT Scala}, except that it uses \systemName DSL instead of standard Scala collections. The performance of the compiled programs for both Scala and C++ output languages are measured.
\end{itemize}

It is worth mentioning that even the baseline is fast and outperforms other IVM techniques \cite{dbt_delta}. The values shown in Figure~\ref{fig:dbt1} are normalized view refresh-rate and not the absolute throughput value. In all cases, we normalized other values with respect to the minimum value.

\begin{table}[htb]
\vspace{4mm}
\caption{Compiler implementation statistics in lines of code.} \label{tab:loc_dbt_compiler}
\vspace{-2mm}
\begin{center}
\small\begin{tabular}{lrrr} \toprule
\bf Component & \bf DBToaster & \bf \systemName & \bf Ratio \\ \midrule
SQL incrementalization${}^{*}$ & 15,874 & -- & -- \\
M3 to Scala compiler  & 14,660 & 1,561 & 11\% \\
\SC${}^{**}$ and inlining${}^{***}$ & 1,891 & 2,713 & 143\% \\
Runtime libraries  & 2,150 & 5,851 & 272\% \\  \bottomrule \vspace{-6pt} \\
\multicolumn{4}{l}{\scriptsize ${}^*$ This is the codebase for converting a SQL query into its corresponding incremental}\\
\multicolumn{4}{l}{\scriptsize ~~~~  view maintenance trigger programs in an intermediate language, named M3.}\\
\multicolumn{4}{l}{\scriptsize ${}^{**}$ Does not include \SC infrastructure (27,433 lines)}\\
\multicolumn{4}{l}{\scriptsize ${}^{***}$ Available only in the new compiler}\\
\end{tabular}\end{center}
\vspace{-2mm}
\end{table}

\paragraph{Results analysis}
From the results in Figure~\ref{fig:dbt1}, we make the following observations: By comparing \{2\} ({\em DBT Scala}) and \{3\} ({\em DBT+\SC Scala}), we see that low-level optimization techniques (DCE, CSE, code motion and limited deforestation) give a small performance improvement to DBToaster programs (at most $1.4\times$ and on average $36\%$ speedup for 22 TPC-H queries).
On the other hand, the programs compiled using \systemName gained more than three orders of magnitude on some queries by applying all the optimizations and an overall $254\times$ average speedup. This much is the difference between only using generic compiler optimizations and staged compilation of database application programs using \systemName.

\begin{figure}
\begin{center}
\leavevmode
\includegraphics[width=\columnwidth]{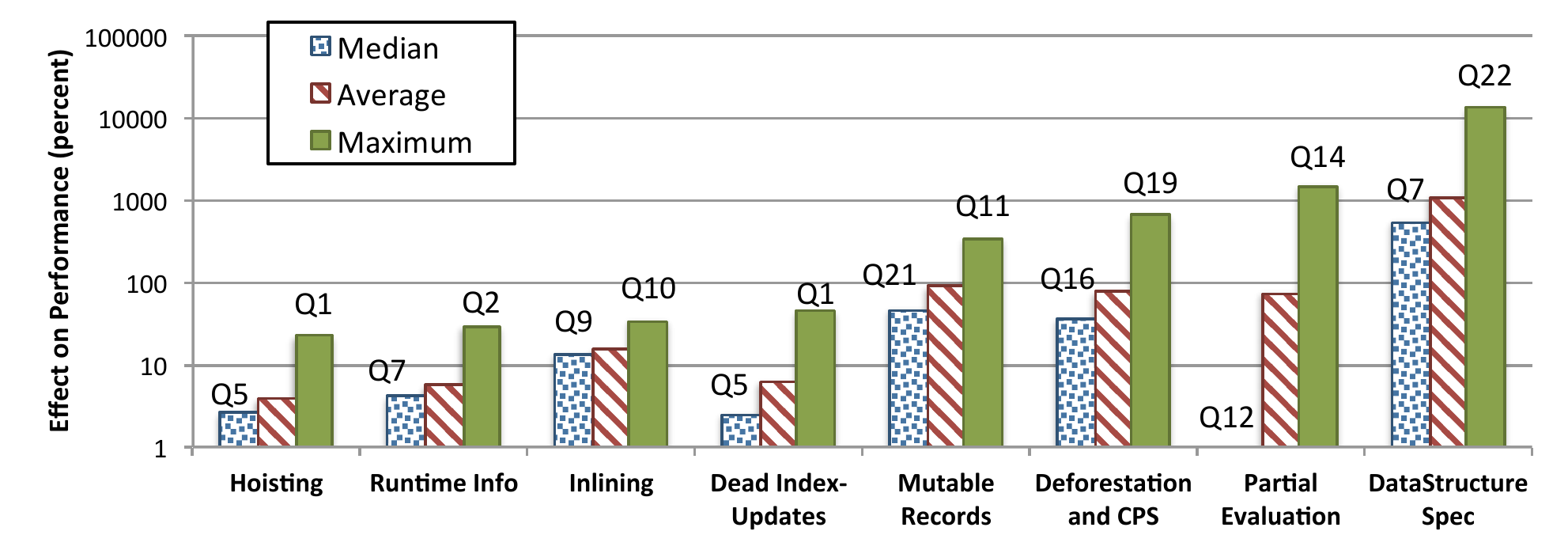}\vspace{-6pt}
\end{center}
\caption{Median, average and maximum impact of important optimizations applied by \systemName to incremental view-maintenance triggers generated by DBToaster for TPC-H queries. The queries with maximum and median impact are shown as data labels on the graph.
}
\label{fig:optImpacts}
\end{figure}

The optimizations in \systemName are modular and can be turned on and off independently  of each other. This allows us to measure the individual impact of each optimization. \{4\} ({\em \systemName C++ -DSS -H}) shows the performance of \systemName in absence of the most influential optimizations and the aggregated break-down of the impact of important optimizations is shown in Figure~\ref{fig:optImpacts}. In this figure, we investigate the median, average and maximum impact of these optimizations. We do not show \textit{automatic index introduction} for generated IVM programs as DBToaster already enables this feature by default.

As shown in Figure~\ref{fig:optImpacts}, {\tt Data Structure Specialization} has the most impact on this benchmark. {\tt Q22} has the largest number of index lookups and field accesses to the data records. Thus, avoiding indirections for these accesses has a tremendous impact on its performance. The impact of {\tt Partial Evaluation} can be attributed to partially evaluating the regular expressions used in the queries and moving them out of the critical execution path. The most significant case is that of {\tt Q13} where there is regular expression matching for each record that is inserted into the {\tt Order} table, which has the highest number of rows in the dataset. In the case of {\tt Dead-Index Updates}, {\tt Q1} shows the maximum improvement owing to its many update operations. Applying the {\tt Mutable Records} optimization impacts {\tt Q11} most, as it has the largest number of index lookups and alterations combined.

Figure~\ref{fig:optImpacts} illustrates that employing the {\tt Runtime Info} does not have a great impact, as the trigger programs mainly use small-sized hash-tables that do not require resizing frequently. Thus, using this optimization, which avoids resizing indexes, does not have a substantial performance improvement. In addition, as hash-indexes are the only type of index used for these trigger programs, index updates are relatively cheap and {\tt Removing Dead Index-Updates} has a moderate impact.

The view maintenance programs for TPC-H queries have many function composition opportunities and thus applying deforestation and {\tt CPS} has a high impact as shown in Figure~\ref{fig:optImpacts}. This optimization has the most impact on {\tt Q19} as it has a chain of seven function compositions in the critical path of the program.

\section{Concurrency Control}\label{sec:concurrency}

In the previous sections, we only focused on achieving the best performance on a single CPU core. In this section, we discuss about the implications of having concurrency control algorithms for \systemName, when a shared-nothing model is not completely satisfactory or applicable.

\subsection{Concurrency Model}

The shared-nothing model with partition-level timestamp ordering, which is the concurrency model used in H-Store \cite{jones10}, is the best match for applying aggressive optimizations to database application programs, as all the programs on a partition are sequentially handled by an isolated worker thread, without any data sharing with other threads. This assumption gives more flexibility to \systemName to apply the optimizations that are only correct in the case of having a single thread that accesses and manipulates the underlying data structures. \systemName can be directly used as a staging compiler for stored procedures at the heart of H-Store.

Even though the H-Store concurrency model performs best for the partitionable workloads \cite{jones10}, not all workloads are perfectly partitionable. The concurrency control algorithms that operate in a shared-memory model (e.g., multi-version concurrency control (MVCC), optimistic concurrency control (OCC), two-phase locking (2PL)), are a better fit for the latter workloads \cite{hekaton,silo}. Using these mechanisms with a shared-memory model has two implications on \systemName: 1) some optimizations are not applicable as-is when these concurrency control algorithms are used, and 2) there might be opportunities for having new optimizations that are tailored for a specific concurrency control algorithm. 

\subsection{Impact on Optimizations}
All the optimizations that we have described for a single-threaded scenario in section \ref{sec:optimizations} are still applicable when multiple threads run on partitioned data. 
However, as using a shared-memory model requires a corresponding change in the underlying data structures, the optimizations that are tied to the data structure have to be redefined. For example, using MVCC implies maintaining several versions for the base data and the indexes are handled accordingly, e.g., by having multi-versioned indexes. In this case, the definition for the {\tt Mutable Records} optimization changes. In a multi-version environment, unlike the single-version scenario, it is necessary to copy-on-write and create a new version upon each update. In order for {\tt Mutable Records} optimization to work in this setting, one naive approach is to have {\tt get} operations that always return a new version of the record on each call. However, this approach is sub-optimal, as not all read operations are followed by an update. Instead, \systemName first performs an analysis to identify the {\tt get} operations that are followed by an {\tt update} and transforms only those {\tt get} operations to {\tt getForUpdate} operations, which returns a newly created version upon each call. Then, any update operation is directly applied on this version, as it is created for this update.

\begin{figure}[t]
\begin{center}
\leavevmode
\includegraphics[width=\columnwidth]{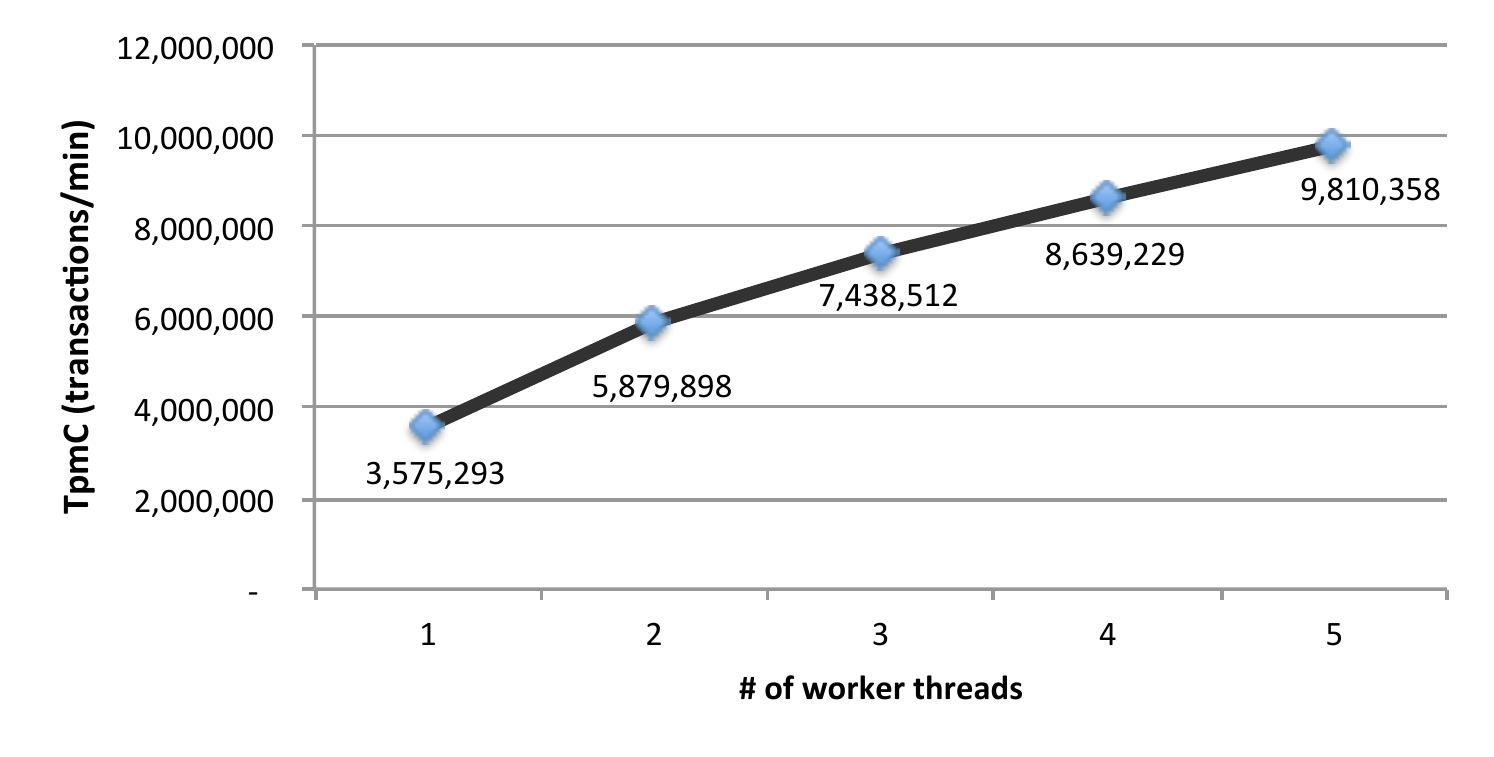}
\end{center}
\vspace{-2mm}
\caption{The scalability of the TPC-C benchmark results using the most optimized generated code by Beta and run under the shared-nothing model (similar to H-Store).}
\vspace{4mm}
\label{fig:tpccHstoreScalability}
\end{figure}

\begin{figure}[htb]
\begin{center}
\leavevmode
	\includegraphics[width=\columnwidth]{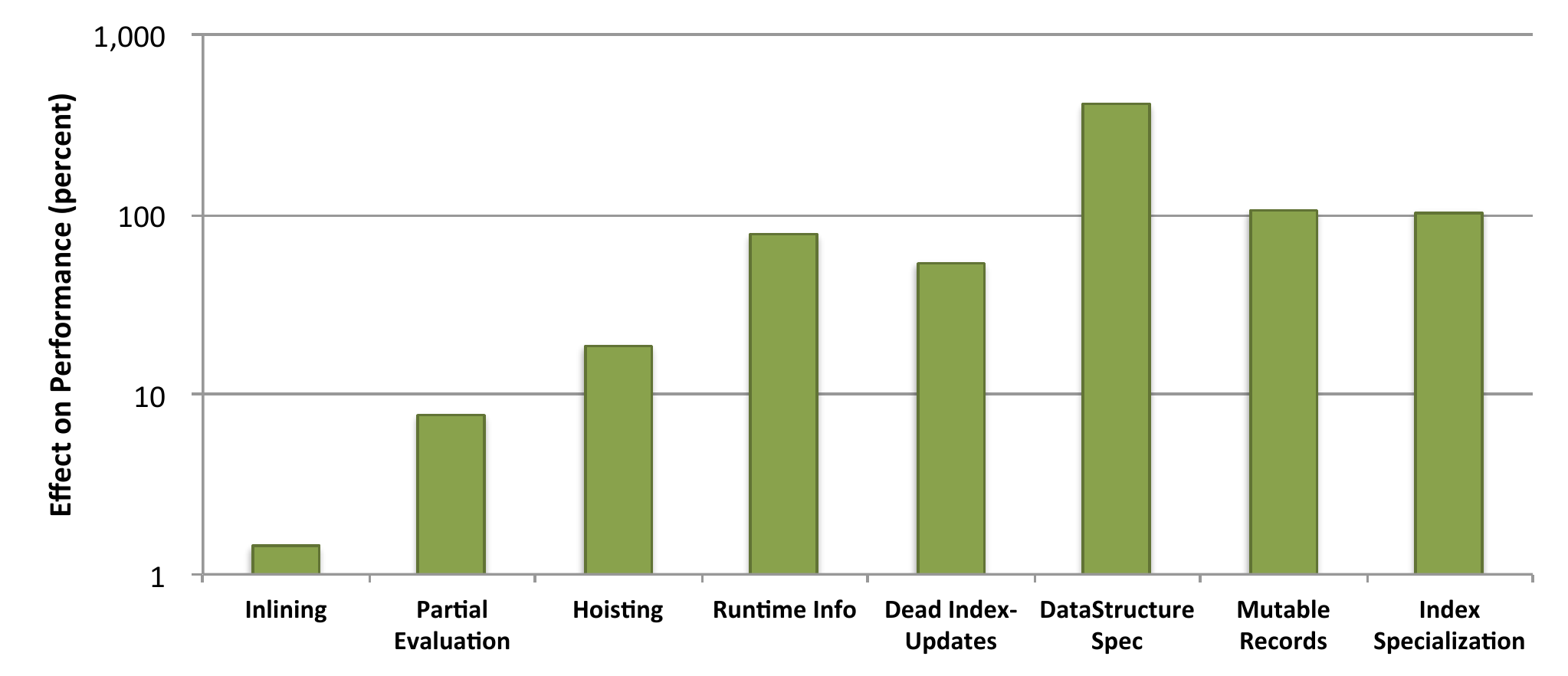}\vspace{-6pt}
\end{center}
\caption{The impact of important optimizations applied by \systemName to TPC-C and ran under the shared-nothing model (similar to H-Store) with five worker threads. 
}
\label{fig:tpccHstoreOptImpacts}
\smallskip
\smallskip
\end{figure}

\subsection{Experiments}
Next, we illustrate the performance of the TPC-C benchmark compiled using \systemName, both when it uses shared-nothing and shared-memory models. These experiments show that using program analysis and compilation techniques still have a significant impact on the performance improvement of database application programs, even in the presence of concurrent and parallel execution.

In order to run the TPC-C benchmark in a shared-nothing model, we took the approach of \cite{hstore_rewrite}. Even though the programs in TPC-C are not perfectly partitionable, it is still possible to divide the programs into smaller pieces, where each piece is executed on one partition, without requiring any synchronization between partitions. This requires a global ordering on the programs executing across partitions, to make sure that the execution is serializable \cite{calvin}.

Figure~\ref{fig:tpccHstoreScalability} shows the overall throughput of TPC-C on the most optimized code generated by \systemName. As expected, the optimizations by \systemName do not hurt the scalability of this benchmark.
In addition, we measured the impact of individual optimizations on the performance of the TPC-C benchmark, when it is run in the shared-nothing model with five worker threads. Again, as expected, the impact of each individual optimization completely matches the one for the single-thread benchmarks presented in section~\ref{sec:evaluation_tpcc}.

\begin{figure}[ht]
\begin{center}
\leavevmode
\includegraphics[width=\columnwidth]{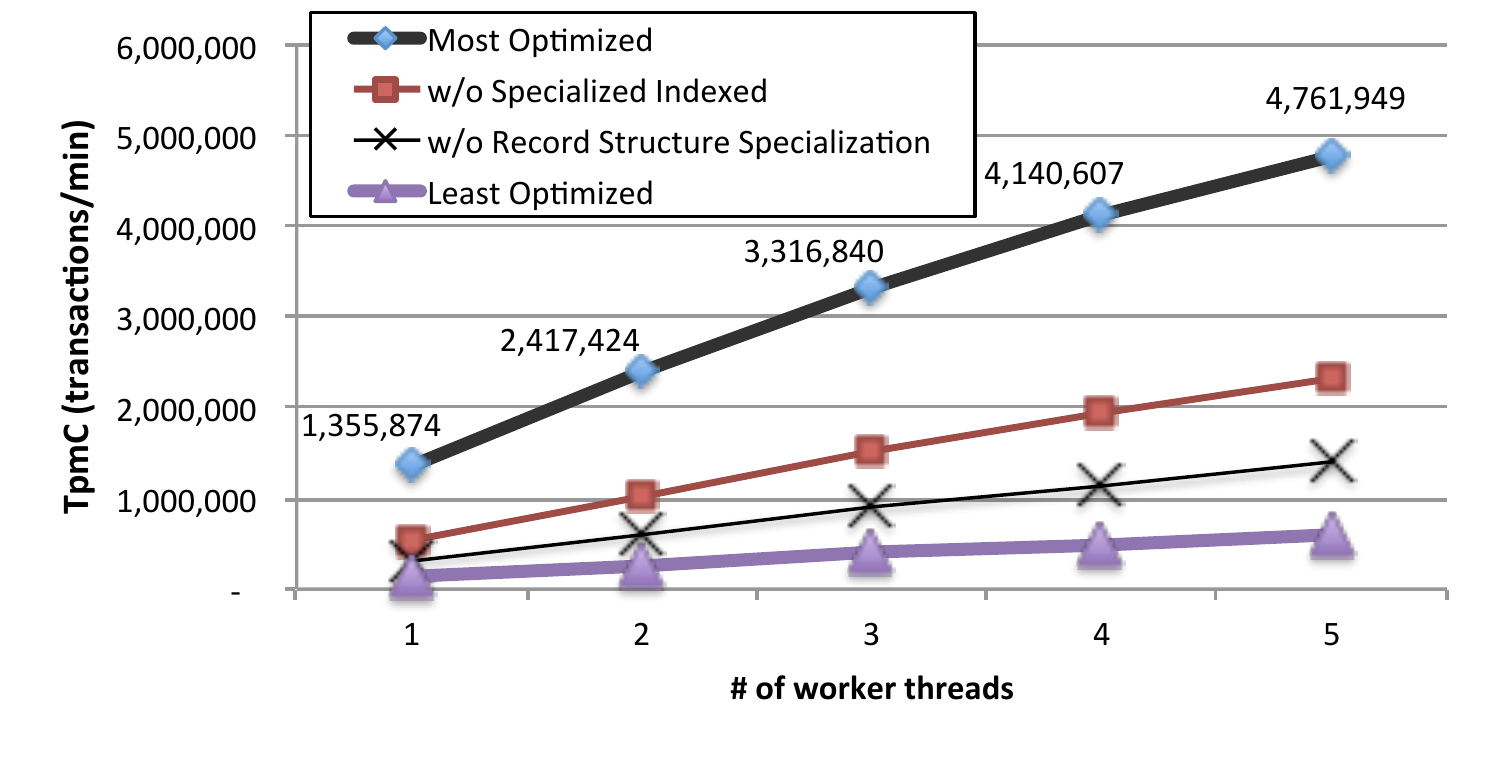}
\end{center}
\caption{The scalability of the TPC-C benchmark results using the generated code by Beta (with different optimizations) and run under optimistic MVCC.}
\label{fig:tpccMvccScalability}
\smallskip
\smallskip
\end{figure}

For the shared-memory model experiments, we chose optimistic MVCC \cite{nmvcc} as the concurrency control algorithm. Then we generated the most optimized programs using \systemName, after applying the necessary modifications to the optimizations for this algorithm. Figure \ref{fig:tpccMvccScalability} shows the throughput for TPC-C in the shared-memory model with different threads for different combination of optimizations. Our code scales linearly with the increase in the number of threads in each of the different optimization combinations, although not at the same rate. The most optimized version not only has the best performance, but also scales the best. The figure also shows that both {\tt Record Structure Specialization} as well as {\tt Specialized Indexes} optimizations play a very important role in achieving good performance in the shared-memory model as well. On the other hand, even though other optimizations such as {\tt Hoisting} and {\tt Partial Evaluation} still optimize the code, their impact is much less compared to the single-threaded model. This is because the performance of the single-threaded model is much better than the base performance (i.e, one thread) of the shared-memory model. As a result, the blocks of code improved by these optimizations form a higher fraction ($f_i$) of the execution time in the generated program (before the optimization) in the single-threaded model  as compared to the shared-memory model, making their impact in the former scenario higher. 

\subsection{Improving Concurrency Control Algorithms}

In this section, we showed the effectiveness of \systemName in presence of using concurrency control algorithms. In addition, there are certain features of the concurrency control algorithms that benefit from the automated program analysis. Consider the case of the attribute-level validation in optimistic MVCC (OMVCC) \cite{nmvcc} instead of the record-level. To do fine-grained attribute-level validation, OMVCC needs to know about the accessed and modified columns for each record. This information can be expensively gathered at runtime, but a better approach is to precompute these columns during compilation. In addition, there are further optimizations that reorder operations to follow the best-practices while using the concurrency control algorithm. As an example, when using 2PL, the access and modification operations of the high-contention records are moved closer to the end of the program in order to reduce the lock lifespan. Furthermore, one can use program dependency information to extend the MVCC algorithm into a more efficient concurrency control algorithm such as MV3C~\cite{mv3c}.

\section{Related work} \label{sec:related}

\paragraph{Front-end DSLs}
The integration of the queries into a programming language, initiated by \cite{linq}, has been adopted by many languages like Scala (Slick) and Ruby.
Recent advances like SWITCH \cite{grust_ruby, Mayr13}, Ferry \cite{grust2010avalanche} and Sloth \cite{sloth} are compiled in the client and produce optimal SQL (with minimal number of executed queries).
Our DSL does not generate queries but compiles application programs directly in the database server and outputs low-level code specialized for the given set of programs. 

\paragraph{Query compilers}
Emerging database systems aim at compiling queries (HyPer \cite{neumann_llvm}, MemSQL, DBLAB/LegoBase \cite{sc,legobase,legobase_tods}) or transaction (H-Store/VoltDB \cite{hstore_rewrite}), but only Hekaton \cite{hekaton} integrates compilation of transactions from a high-level language (SQL). The main differences with \systemName are:\begin{itemize}
\item {\em Compilation}: \systemName uses a unified DSL with two abstraction levels, Hekaton uses two disjoint internal representations, i.e., mixed abstract tree (MAT) and pure imperative tree (PIT).
Hekaton inlines and wires iterator functions using {\tt goto}s, this still keeps some intermediate materializations; with fusion and CPS, we get rid of them.
\item {\em Data structures}: Hekaton supports predefined indexes (Bw tree, hash with fixed number of buckets); \systemName chooses the most appropriate dynamic data structure based on a restricted set of queries.
\item {\em Environment}: Hekaton fixes the concurrency model to MVCC, but \systemName handles multiple concurrency models (see section \ref{sec:concurrency}). Hekaton compiles directly to C whereas \systemName relies on multiple optimization layers and as the last stage generates C code.
\end{itemize}

Hyper \cite{neumann_llvm} generates low-level code using LLVM to improve query performance. The system is different from \systemName in that Hyper only focuses on individual transaction programs and does not even compile whole transaction programs, but individual queries. Furthermore, \systemName leverages its modular design to give insight into the performance impact of each individual optimization (as shown in the experiments), while such a facility is not reported for Hyper. 

\paragraph{Application-level compilation}
Database applications are commonly implemented using a combination of procedural and 
declarative languages. The business logic is usually implemented in an imperative language such as C\# or Java, the data access layer is programmed using SQL, and an application programming interface (API) such as Java Database Connectivity (JDBC) is used to communicate the SQL queries between the Java program and the database. Having these two environments can introduce a downgrade in performance. The main reason is that neither of these environments can make use of the optimization opportunities in the other side.
 For example, the database indexes are not known by the Java environment and the database system is not aware of the loops in the Java code~\cite{unifiedadhoc}.

One way to optimize such programs is by using program analysis techniques to extract declarative queries from the imperative code~\cite{Cheung:2013:ODA:2491956.2462180, cheung2014using, cheung2013statusquo, Wiedermann:2007:EQS:1190216.1190248, Wiedermann:2008:IQE:1449764.1449767}. As a result, the extracted code can benefit from the optimizations offered by the underlying database system. Furthermore, it is possible to partition database applications between the application runtime and the database system~\cite{cheung2013statusquo,Cheung:2012:APD:2350229.2350262}, merge several related queries into a single query~\cite{Guravannavar:2008:RPB:1453856.1453975,manjhi2009holistic}, and prefetch the query results~\cite{ramachandra2012holistic,chavan2011program}. However, as SQL is not as expressive as an imperative language, this approach is not applicable to all database applications. In addition, for applying optimizations available at a lower level of abstraction (e.g. operator inlining, inter-operator optimization, etc.), one should rely on the database system.

The alternative approach is to rewrite both the application logic as well as the data access part into an intermediate language, such as UniQL~\cite{unifiedadhoc} and forelem~\cite{rietveld2015reducing}. This way, all the optimizations happening in \textit{both} the application runtime (e.g. the underlying optimization compiler of the application program) and the database system (e.g. query optimization) become applicable directly. Although the intermediate language in such systems is expressive enough, these systems mainly focus on high-level optimizations available at the corresponding intermediate language~\cite{unifiedadhoc}. In contrast, our approach utilizes multiple levels of abstraction and, thus, makes it possible to perform optimizations available across different abstraction levels, such as {\tt Record Structure Specialization}.

Recently, there were several efforts in order to boost the performance 
of the database applications written using \textit{language integrated queries} (LINQ~\cite{linq})  using database-inspired strategies and optimizations through code generation and just-in-time compilation~\cite{ferry-2,Murray:2011:SAO:1993498.1993513,Nagel:2014:CGE:2732977.2732984,DBLP:journals/debu/ViglasBN14}. 
In general, all these techniques employ compilation to convert high-level LINQ programs to more efficient, imperative low-level code. 
This line of work mostly targets making query processing of collections in the memory space of the 
application more efficient by leveraging database technology. 
As an example, \cite{DBLP:journals/debu/ViglasBN14} can modify the memory layout of a collection of records, from a generic array of 
pointers to objects allocated on the managed heap, into an array of contiguous objects. However, due to the lack of multiple intermediate 
languages in these systems, it is not possible to support {\tt Record Structure Specialization}, {\tt Mutable Records}, etc. 

\balance
\bibliographystyle{abbrv}
\bibliography{refs}

\begin{thebibliography}{10}

\bibitem{dbt_delta}
Y.~Ahmad, O.~Kennedy, C.~Koch, and M.~Nikolic.
\newblock {DBToaster}: Higher-order delta processing for dynamic, frequently
  fresh views.
\newblock {\em PVLDB}, 5(10):968--979, June 2012.

\bibitem{appel_cps}
A.~W. Appel and T.~Jim.
\newblock Continuation-passing, closure-passing style.
\newblock In {\em POPL}, pages 293--302, 1989.

\bibitem{monad-calc-1}
V.~Breazu-Tannen, P.~Buneman, and L.~Wong.
\newblock {\em Naturally embedded query languages}.
\newblock Springer, 1992.

\bibitem{meta_ocaml}
C.~Calcagno, W.~Taha, L.~Huang, and X.~Leroy.
\newblock Implementing multi-stage languages using {ASTs}, {Gensym}, and
  reflection.
\newblock In {\em GPCE}, 2003.

\bibitem{lms_hp1}
H.~Chafi, Z.~DeVito, A.~Moors, T.~Rompf, A.~K. Sujeeth, P.~Hanrahan,
  M.~Odersky, and K.~Olukotun.
\newblock Language virtualization for heterogeneous parallel computing.
\newblock In {\em OOPSLA}, pages 835--847, 2010.

\bibitem{DBLP:journals/cacm/ChamberlinABGKLLMPPSSSTWY81}
D.~D. Chamberlin, M.~M. Astrahan, M.~W. Blasgen, J.~Gray, W.~F.~K. III, B.~G.
  Lindsay, R.~A. Lorie, J.~W. Mehl, T.~G. Price, G.~R. Putzolu, P.~G. Selinger,
  M.~Schkolnick, D.~R. Slutz, I.~L. Traiger, B.~W. Wade, and R.~A. Yost.
\newblock A history and evaluation of {System R}.
\newblock {\em Commun. ACM}, 24(10):632--646, 1981.

\bibitem{indexes_olap2}
S.~Chaudhuri, M.~Datar, and V.~Narasayya.
\newblock Index selection for databases: a hardness study and a principled
  heuristic solution.
\newblock {\em IEEE Trans. Knowl. Data Eng.}, 2004.

\bibitem{indexes_tool}
S.~Chaudhuri and V.~Narasayya.
\newblock An efficient, cost-driven index selection tool for microsoft sql
  server.
\newblock In {\em VLDB}, volume~97, pages 146--155, 1997.

\bibitem{chavan2011program}
M.~Chavan, R.~Guravannavar, K.~Ramachandra, and S.~Sudarshan.
\newblock Program transformations for asynchronous query submission.
\newblock ICDE'11, pages 375--386. IEEE, 2011.

\bibitem{cheung2013statusquo}
A.~Cheung, O.~Arden, S.~Madden, A.~Solar-Lezama, and A.~C. Myers.
\newblock Statusquo: Making familiar abstractions perform using program
  analysis.
\newblock In {\em CIDR}. www.cidrdb.org, 2013.

\bibitem{Cheung:2012:APD:2350229.2350262}
A.~Cheung, S.~Madden, O.~Arden, and A.~C. Myers.
\newblock Automatic partitioning of database applications.
\newblock {\em Proc. VLDB Endow.}, 5(11):1471--1482, July 2012.

\bibitem{sloth}
A.~Cheung, S.~Madden, and A.~Solar-Lezama.
\newblock Sloth: Being lazy is a virtue (when issuing database queries).
\newblock In {\em Proceedings of the 2014 ACM SIGMOD International Conference
  on Management of Data}, SIGMOD '14, pages 931--942, New York, NY, USA, 2014.
  ACM.

\bibitem{cheung2014using}
A.~Cheung, S.~Madden, A.~Solar-Lezama, O.~Arden, and A.~C. Myers.
\newblock Using program analysis to improve database applications.
\newblock {\em IEEE Data Eng. Bull.}, 37(1):48--59, 2014.

\bibitem{Cheung:2013:ODA:2491956.2462180}
A.~Cheung, A.~Solar-Lezama, and S.~Madden.
\newblock Optimizing database-backed applications with query synthesis.
\newblock PLDI '13, pages 3--14, New York, NY, USA, 2013. ACM.

\bibitem{mv3c}
M.~Dashti, S.~B. John, A.~Shaikhha, and C.~Koch.
\newblock Transaction repair for multi-version concurrency control.
\newblock In {\em SIGMOD}, 2017.

\bibitem{hekaton}
C.~Diaconu, C.~Freedman, E.~Ismert, P.-A. Larson, P.~Mittal, R.~Stonecipher,
  N.~Verma, and M.~Zwilling.
\newblock Hekaton: {SQL Server}'s memory-optimized {OLTP} engine.
\newblock In {\em SIGMOD}, 2013.

\bibitem{duboscq2013graal}
G.~Duboscq, L.~Stadler, T.~W{\"u}rthinger, D.~Simon, C.~Wimmer, and
  H.~M{\"o}ssenb{\"o}ck.
\newblock Graal ir: An extensible declarative intermediate representation.
\newblock In {\em Proceedings of the Asia-Pacific Programming Languages and
  Compilers Workshop}, 2013.

\bibitem{compile_edsl}
C.~Elliott, S.~Finne, and O.~De~Moor.
\newblock Compiling embedded languages.
\newblock {\em J. Funct. Program.}, pages 455--481, May 2003.

\bibitem{foldr-fusion-1}
A.~Gill, J.~Launchbury, and S.~L. Peyton~Jones.
\newblock A short cut to deforestation.
\newblock FPCA, pages 223--232. ACM, 1993.

\bibitem{grust_ruby}
T.~Grust and M.~Mayr.
\newblock A deep embedding of queries into {Ruby}.
\newblock In {\em ICDE}, 2012.

\bibitem{ferry-2}
T.~Grust, M.~Mayr, J.~Rittinger, and T.~Schreiber.
\newblock Ferry -- database-supported program execution.
\newblock SIGMOD '09, pages 1063--1066, New York, NY, USA, 2009. ACM.

\bibitem{grust2010avalanche}
T.~Grust, J.~Rittinger, and T.~Schreiber.
\newblock Avalanche-safe linq compilation.
\newblock {\em Proceedings of the VLDB Endowment}, 3(1-2):162--172, 2010.

\bibitem{indexes_olap}
H.~Gupta, V.~Harinarayan, A.~Rajaraman, and J.~D. Ullman.
\newblock Index selection for {OLAP}.
\newblock In {\em ICDE}, 1997.

\bibitem{Guravannavar:2008:RPB:1453856.1453975}
R.~Guravannavar and S.~Sudarshan.
\newblock Rewriting procedures for batched bindings.
\newblock {\em Proc. VLDB Endow.}, 1(1):1107--1123, Aug. 2008.

\bibitem{contdelivery}
J.~Humble and D.~Farley.
\newblock {\em Continuous Delivery: Reliable Software Releases Through Build,
  Test, and Deployment Automation}.
\newblock Addison-Wesley Professional, 1st edition, 2010.

\bibitem{jones10}
E.~P. Jones, D.~J. Abadi, and S.~Madden.
\newblock Low overhead concurrency control for partitioned main memory
  databases.
\newblock In {\em Proceedings of the 2010 international conference on
  Management of data, SIGMOD '10}, pages 603--614, New York, NY, USA, 2010.
  ACM.

\bibitem{jones1993partial}
N.~D. Jones, C.~K. Gomard, and P.~Sestoft.
\newblock {\em Partial evaluation and automatic program generation}.
\newblock Peter Sestoft, 1993.

\bibitem{haskell_nest_para}
S.~P. Jones, R.~Leshchinskiy, G.~Keller, and M.~M.~T. Chakravarty.
\newblock Harnessing the multicores: Nested data parallelism in {Haskell}.
\newblock In {\em FSTTCS}, pages 383--414, 2008.

\bibitem{dbt_agile}
O.~Kennedy, Y.~Ahmad, and C.~Koch.
\newblock {DBToaster}: Agile views for a dynamic data management system.
\newblock In {\em CIDR}, pages 284--295, 2011.

\bibitem{legobase}
Y.~Klonatos, C.~Koch, T.~Rompf, and H.~Chafi.
\newblock Building efficient query engines in a high-level language.
\newblock {\em Proc. VLDB Endow.}, 7(10):853--864, June 2014.

\bibitem{KochManifesto}
C.~Koch.
\newblock Abstraction without regret in database systems building: a manifesto.
\newblock {\em IEEE Data Eng.\ Bull.}, 37(1), 2014.

\bibitem{manjhi2009holistic}
A.~Manjhi, C.~Garrod, B.~M. Maggs, T.~C. Mowry, and A.~Tomasic.
\newblock Holistic query transformations for dynamic web applications.
\newblock In {\em Proceedings of the 2009 IEEE International Conference on Data
  Engineering}, ICDE '09, pages 1175--1178, Washington, DC, USA, 2009. IEEE
  Computer Society.

\bibitem{Mayr13}
M.~Mayr.
\newblock {\em A Deep Embedding of Queries into Ruby}.
\newblock PhD thesis, Universit\"at T\"ubingen, 2013.

\bibitem{linq}
E.~Meijer, B.~Beckman, and G.~Bierman.
\newblock Linq: Reconciling object, relations and xml in the .net framework.
\newblock In {\em SIGMOD}, pages 706--706, 2006.

\bibitem{advancedcom}
S.~S. Muchnick.
\newblock {\em Advanced Compiler Design and Implementation}.
\newblock Morgan Kaufmann Publishers Inc., San Francisco, CA, USA, 1997.

\bibitem{Murray:2011:SAO:1993498.1993513}
D.~G. Murray, M.~Isard, and Y.~Yu.
\newblock {Steno: Automatic Optimization of Declarative Queries}.
\newblock PLDI '11, pages 121--131, New York, NY, USA, 2011. ACM.

\bibitem{Nagel:2014:CGE:2732977.2732984}
F.~Nagel, G.~Bierman, and S.~D. Viglas.
\newblock Code generation for efficient query processing in managed runtimes.
\newblock {\em Proc. VLDB Endow.}, 7(12):1095--1106, Aug. 2014.

\bibitem{neumann_llvm}
T.~Neumann.
\newblock Efficiently compiling efficient query plans for modern hardware.
\newblock {\em PVLDB}, 4(9), June 2011.

\bibitem{nmvcc}
T.~Neumann, T.~M\"{u}hlbauer, and A.~Kemper.
\newblock Fast serializable multi-version concurrency control for main-memory
  database systems.
\newblock In {\em SIGMOD}, 2015.

\bibitem{scala_overview}
M.~Odersky, P.~Altherr, V.~Cremet, B.~Emir, S.~Maneth, S.~Micheloud,
  N.~Mihaylov, M.~Schinz, E.~Stenman, and M.~Zenger.
\newblock An overview of the {Scala} programming language.
\newblock Technical report, EPFL, 2004.

\bibitem{halide}
J.~Ragan-Kelley, C.~Barnes, A.~Adams, S.~Paris, F.~Durand, and S.~Amarasinghe.
\newblock Halide: A language and compiler for optimizing parallelism, locality,
  and recomputation in image processing pipelines.
\newblock {\em SIGPLAN Not.}, 48(6):519--530, June 2013.

\bibitem{ramachandra2012holistic}
K.~Ramachandra and S.~Sudarshan.
\newblock Holistic optimization by prefetching query results.
\newblock In {\em Proceedings of the 2012 ACM SIGMOD International Conference
  on Management of Data}, pages 133--144. ACM, 2012.

\bibitem{rietveld2015reducing}
K.~F.~D. Rietveld and H.~A.~G. Wijshoff.
\newblock Reducing layered database applications to their essence through
  vertical integration.
\newblock {\em ACM Trans. Database Syst.}, 40(3):18:1--18:39, Oct. 2015.

\bibitem{lms_original}
T.~Rompf and M.~Odersky.
\newblock Lightweight modular staging: A pragmatic approach to runtime code
  generation and compiled {DSLs}.
\newblock In {\em GPCE}, pages 127--136, 2010.

\bibitem{lms_opt_datastruct}
T.~Rompf, A.~K. Sujeeth, N.~Amin, K.~J. Brown, V.~Jovanovic, H.~Lee,
  M.~Jonnalagedda, K.~Olukotun, and M.~Odersky.
\newblock Optimizing data structures in high-level programs: New directions for
  extensible compilers based on staging.
\newblock In {\em POPL}, pages 497--510, 2013.

\bibitem{plprag}
M.~L. Scott.
\newblock {\em Programming Language Pragmatics, Third Edition}.
\newblock Morgan Kaufmann Publishers Inc., San Francisco, CA, USA, 3rd edition,
  2009.

\bibitem{jfppushpull}
A.~Shaikhha, M.~Dashti, and C.~Koch.
\newblock {Push versus Pull-Based Loop Fusion in Query Engines}.
\newblock {\em Journal of Functional Programming}, 28:e10, 2018.

\bibitem{legobase_tods}
A.~Shaikhha, Y.~Klonatos, and C.~Koch.
\newblock Building efficient query engines in a high-level language.
\newblock {\em ACM Transactions on Database Systems}, 43(1):4:1--4:45, Apr.
  2018.

\bibitem{sc}
A.~Shaikhha, Y.~Klonatos, L.~Parreaux, L.~Brown, M.~Dashti, and C.~Koch.
\newblock How to architect a query compiler.
\newblock In {\em SIGMOD}, 2016.

\bibitem{unifiedadhoc}
X.~Shi, B.~Cui, G.~Dobbie, and B.~C. Ooi.
\newblock Towards unified ad-hoc data processing.
\newblock In {\em Proceedings of the 2014 ACM SIGMOD International Conference
  on Management of Data}, SIGMOD '14, pages 1263--1274, New York, NY, USA,
  2014. ACM.

\bibitem{old_opt_relalg}
J.~M. Smith and P.~Y.-T. Chang.
\newblock Optimizing the performance of a relational algebra database
  interface.
\newblock {\em Commun. ACM}, 18(10):568--579, 1975.

\bibitem{hstore_rewrite}
M.~Stonebraker, S.~Madden, D.~J. Abadi, S.~Harizopoulos, N.~Hachem, and
  P.~Helland.
\newblock The end of an architectural era: (it's time for a complete rewrite).
\newblock In {\em VLDB}, pages 1150--1160, 2007.

\bibitem{delitejournal}
A.~K. Sujeeth, K.~J. Brown, H.~Lee, T.~Rompf, H.~Chafi, M.~Odersky, and
  K.~Olukotun.
\newblock Delite: A compiler architecture for performance-oriented embedded
  domain-specific languages.
\newblock {\em ACM Trans. Embed. Comput. Syst.}, 13(4s):134:1--134:25, Apr.
  2014.

\bibitem{optiml}
A.~K. Sujeeth, H.~Lee, K.~J. Brown, H.~Chafi, M.~Wu, A.~R. Atreya, K.~Olukotun,
  T.~Rompf, and M.~Odersky.
\newblock {OptiML}: an implicitly parallel domain-specific language for machine
  learning.
\newblock In {\em ICML}, 2011.

\bibitem{calvin}
A.~Thomson, T.~Diamond, S.-C. Weng, K.~Ren, P.~Shao, and D.~J. Abadi.
\newblock Calvin: Fast distributed transactions for partitioned database
  systems.
\newblock In {\em Proceedings of the 2012 ACM SIGMOD International Conference
  on Management of Data}, SIGMOD '12, pages 1--12, New York, NY, USA, 2012.
  ACM.

\bibitem{silo}
S.~Tu, W.~Zheng, E.~Kohler, B.~Liskov, and S.~Madden.
\newblock Speedy transactions in multicore in-memory databases.
\newblock In {\em Proceedings of the Twenty-Fourth ACM Symposium on Operating
  Systems Principles}, SOSP '13, pages 18--32, New York, NY, USA, 2013. ACM.

\bibitem{DBLP:journals/debu/ViglasBN14}
S.~Viglas, G.~M. Bierman, and F.~Nagel.
\newblock Processing\hspace{0.05cm} declarative\hspace{0.05cm}
  queries\hspace{0.05cm} through\hspace{0.05cm} generating\hspace{0.05cm}
  imperative\hspace{0.05cm} code\hspace{0.05cm} in\hspace{0.05cm}
  managed\hspace{0.05cm} runtimes.
\newblock {\em {IEEE} Data Eng. Bull.}, 37(1):12--21, 2014.

\bibitem{spiral}
Y.~Voronenko, F.~Franchetti, F.~{de Mesmay}, and M.~P{\"u}schel.
\newblock Generating high-performance general size linear transform libraries
  using {Spiral}.
\newblock In {\em HPEC}, 2008.

\bibitem{deforestation}
P.~Wadler.
\newblock Deforestation: Transforming programs to eliminate trees.
\newblock In {\em ESOP'88}, pages 344--358. Springer, 1988.

\bibitem{Wiedermann:2007:EQS:1190216.1190248}
B.~Wiedermann and W.~R. Cook.
\newblock Extracting queries by static analysis of transparent persistence.
\newblock POPL '07, pages 199--210, New York, NY, USA, 2007. ACM.

\bibitem{Wiedermann:2008:IQE:1449764.1449767}
B.~Wiedermann, A.~Ibrahim, and W.~R. Cook.
\newblock Interprocedural query extraction for transparent persistence.
\newblock OOPSLA '08, pages 19--36, New York, NY, USA, 2008. ACM.

\end{thebibliography}

\end{document}